\documentclass[a4,11pt]{article}

\usepackage{amsmath}
\usepackage{amsfonts}
\usepackage{amssymb}
\usepackage{latexsym}
\usepackage{multicol}
\usepackage{color}
\usepackage[utf8]{inputenc}
\usepackage{graphicx}
\usepackage{dsfont}
\usepackage[T1]{fontenc}
\usepackage{natbib}
\usepackage{multirow}

\def \beg{\begin{eqnarray}}
\def \en{\end{eqnarray}}
\def \be*{\begin{eqnarray*}}
\def\e*{\end{eqnarray*}}
\def \di{\displaystyle}
\def\bit{\begin{itemize}}
\def \eit{\end{itemize}}

\def \cqfd{\hspace*{13cm}$\Box$}
\def \tpi{\tilde{\pi}}

\newtheorem{prp}{Proposition}[section]
\newtheorem{rmk}{Remark}[section]
\newtheorem{lmm}{Lemma}[section]

\begin{document}

\begin{center}
{\Large
	{\sc
	 Parallel Tempering with Equi-Energy Moves
	}
}
\bigskip

Meïli Baragatti$^{1,2,*}$, Agnès Grimaud$^{2}$, Denys Pommeret$^{2}$

\medskip
{\it
 $^1$ Ipsogen SA, Luminy Biotech Entreprises, Case 923, Campus de Luminy, 13288 Marseille Cedex 9, France.\\
 $^2$ Institut de Mathématiques de Luminy (IML), CNRS Marseille, case 907, Campus de Luminy, 13288 Marseille Cedex 9, France.\\
 $^*$ baragatt@iml.univ-mrs.fr, baragattimeili@hotmail.com.
}
\end{center}

\begin{center}
{\sc Working Paper\\ 5th October 2011}
\end{center}

\bigskip
\noindent

\begin{abstract}
The Equi-Energy Sampler (EES) introduced by \cite{KouZhouWong} is based on a population of chains which are updated by local moves and global moves, also called equi-energy jumps. The state space is partitioned into energy rings, and the current state of a chain can jump to a past state of an adjacent chain that has an energy level close to its level. This algorithm has been developed to facilitate global moves between different chains, resulting in a good exploration of the state space by the target chain.
This method seems to be more efficient than the classical Parallel Tempering (PT) algorithm.
However it is difficult to use in combination with a Gibbs sampler and it necessitates increased storage.
In this paper we propose an adaptation of this EES that combines PT with the principle of swapping between chains with same levels of energy. This adaptation, that we shall call Parallel Tempering with Equi-Energy Moves (PTEEM),  keeps the original idea of the EES method while ensuring good theoretical properties, and practical implementation. 
Performances of the PTEEM algorithm are compared with those of the EES and of the standard PT algorithms in the context of mixture models, and in a problem of identification of transcription factor binding motifs.
\end{abstract}

\section{Introduction}
      A common  problem in Bayesian statistics  is that of generating random variables from
      a target density $\pi$. Many solutions have been proposed in the last two decades, deriving essentially from the Monte Carlo Markov Chains (MCMC) approach introduced by \cite{Metropolis1953} and \cite{Hastings1970}. In classical MCMC methods, a Markov process is built to sample the target probability distribution.
      But in practice, the Markov process can be easily trapped into a local mode from where it cannot escape in reasonable time (see for instance \cite{LiangWong}).
      Many techniques have been proposed to address this waiting time problem, including among others  Parallel Tempering (PT) (see \cite{Geyer1991} or \cite{GeyerThompson}), and more recently the Equi-Energy Sampler (EES) (\cite{KouZhouWong}).
%      On some state space $\mathcal X$ with associated $\sigma$-algebra $\cal B (\cal X)$,  the target density is denoted by $\pi(x)  \propto  \exp\{-h(x)\}$, where $h(x) $ denotes the energy function.

      In the PT algorithm, $N$ temperatures are introduced, and $N$ chains are run in parallel, with target distributions being tempered distributions of the target $\pi$. Note that the first chain targets $\pi$. Since the tempered distributions becomes flatter as the temperature increases, the chains at high temperatures can move easily
      between modes. Each iteration of the PT algorithm is decomposed into two types of moves: local moves via classical MCMC algorithms to update the different chains, and global moves allowing swaps between two chains. The use of these swaps enables new modes to be propagated through the different chains, thereby improving mixing. The first chain associated with the target distribution will then be able to escape from local modes.
%       The principle of PT is to choose $N$ temperatures and to run in parallel $N$ associated MCMC chains having related
%       stationary distributions, the first chain targeting the distribution of interest. Since the tempered distribution becomes flatter as the temperature increases, the chains at high temperatures can move easily
%      The PT algorithm consists of two steps at each iteration: a parallel step updating every chain by using their respective MCMC algorithm, and a swapping step consisting in choosing randomly two chains and in proposing a swap between those.
%       An advantage of this algorithm is its ability to use information from different chains through the swapping step. Therefore the swapping step allows the chain associated with the posterior distribution to escape from its local modes, thereby improving mixing.
      Some improvements of PT have been proposed, like swaps with delayed rejection (see \cite{GreenMira}) which permit to propose a new swap when the first one is not accepted, or like Evolutionary Monte Carlo (\cite{LiangWong}). However this PT algorithm does not retain information of where chains have been and it does not choose one of the best swaps. This is what is done by the EES proposed by \cite{KouZhouWong}, by using a partition of the state space along the energy function. Note that such a partitioning has already been recommended by \cite{Mitsutake2003} and \cite{AtchadeLiu2006}, in an importance sampling framework.

      In the EES algorithm,  the target density is rewritten in terms of  energy function, and $K$ temperatures and energy levels are introduced. Then a population of $K$ distributions is considered, each one being a tempered distribution of $\pi$ truncated by an energy level. This algorithm is mainly based on a new type of move called the equi-energy jump, that aims to explore the state space by moving directly between states with similar energy. The goal is still to improve mixing of the chains. However, to perform these moves, the sampler uses past states of the different chains. All these past states should then be kept in memory.
      A substantial advantage of this algorithm is that it seems to be very efficient compared to classical MCMC methods and to PT (see \cite{KouZhouWong}).
      But an associated drawback is the cost of increased storage, all the past being taken into account in equi-energy jumps. In addition some difficulties are encountered to combine EES with a  Gibbs sampler. The problem is to sample from the tempered distributions truncated by energy levels. Some algorithms could be used to sample from it, like accept-reject or Approximate Bayesian Computation algorithms, but the computational cost would then be too high in practice.
      From a theoretical point of view, the EES is not based on a Markov chain, and its theoretical analysis is relatively difficult. Several authors studied its convergence under various assumptions. The proof of the convergence has been discussed in \cite{AtchadeLiu2006}, \cite{Andrieu2007}, \cite{AndrieuJasra2007} and \cite{Andrieu2008}. \cite{HuaKou} completed the proof of the convergence of the EES in the case of a countable state space, and recently more general convergence results has been established by \cite{AtchadeFortMoulinesPriouret}. Note that \cite{Atchade2010b} showed that the asymptotic variances of adaptive MCMC algorithms (and hence the EES) are always at least as large as the asymptotic variances of MCMC algorithms with the same target distributions, and that the differences can be substantial.

      In this paper we develop an adaptation of the PT and EES algorithms,  called the Parallel Tempering with Equi-Energy Moves (PTEEM) algorithm. An equi-energy exchange move is proposed, based on the energies of current states of the chains, and not on past states. Compared to PT algorithm, only moves between chains whose states are close in energy are proposed. This focuses computational effort on moves which are likely to be accepted, and hence which allow jumps between modes. This PTEEM algorithm can be easily combined with a Gibbs sampler, and its convergence is ensured. Furthermore, it does not need a large storage.
 The possible loss or gain of this algorithm compared to EES  and PT are evaluated through simulations and real data. %The advantage of using a Gibbs sampling is highlighted on a problem of  motif sampling already studied by \cite{KouZhouWong}.

%       We propose to adapt the concept of equi-energy jump in a schema of PT, with rings of energy depending only on the currents chains, not on all the past. Then the jump between chains still depends on their energies, and the candidate chains for swapping are chosen randomly and uniformly. This combination of EES and PT yields to the new algorithm PTEEM for which convergence is ensured (see Propositions \ref{prop1} and \ref{prop2}). Comparing to PT algorithm, the proposed method only proposes moves between chains whose states are close in energy. This focuses computational effort on the moves that are likely to lead to jumps between modes. Furthermore, unlike in the  EES schema,  PTEEM algorithm can be combined with a Gibbs sampling (see Remark \ref{rmk1}).
%       Another significant advantage of the proposed method relative to the EES is that, since it is a Markov chain, it does not require a large and increasing storage.
%       The possible loss or gain of this algorithm compared to EES  and PT are evaluated through simulations and real data. The advantage of using a Gibbs sampling is highlighted on a problem of  motif sampling already studied by Kou {\it et al.} \cite{KouZhouWong}.

      The paper is organized as follows:
      In Section 2, PT and EES algorithms are briefly recalled. In Section 3 the PTEEM algorithm is presented.
      In Sections 4 and 5, performances of the PTEEM algorithm are compared with those of the EES and of the standard PT algorithms in the context of mixture models, through  simulations and real data. In Section 6, PTEEM and EES algorithms are compared in a challenging problem of identification of transcription factor binding motifs.  Section 7 presents concluding remarks.

\section{Background on PT and EES algorithms}
      \subsection{PT algorithm}
% 	    On some state space $\mathcal X$ with associated $\sigma$-algebra $\cal B (\cal X)$,  the target density is proportional to
% 	    \be*
% 	    \pi(x) & \propto & \exp\{-h(x)\},
% 	    \e*
% 	    where $h(x) $ denotes the energy function.
% 	    In a classical Metropolis-Hasting algorithm a
% 	    new state $y$ is generated from a current state $x$ of the Markov process by drawing $y$ from a proposal transition function
% 	    $q(x;y)$. The new state $y$ is accepted with the probability $\min(1,r)$, where $r$ is the Metropolis-Hastings ratio:
% 	    \be*
% 	    r &= & \di\frac{\pi(y)q(y;x)}{\pi(x)q(x;y)}.
% 	    \e*
% 	    The Markov process converges to the target distribution $\pi$ using any positive transition function
% 	    $q(x;y)$ and starting from any initial configuration. Nevertheless, in practice, the Markov process can be trapped into a deep local
% 	    minimum of energy.
	    In case of complex or high dimensional problems whose densities of interest contain several modes, classical MCMC methods (like Metropolis-Hastings algorithm or Gibbs sampler for instance) are often trapped into local modes from where they cannot escape in reasonable time. To avoid this problem, the principle of PT is to choose $N$ temperatures $T_1=1 < T_2 < \cdots < T_N$, and to run in parallel $N$ associated MCMC chains having different
	    stationary distributions,  $\pi_1, \cdots, \pi_N$, where
	    \be*
	    \pi_i& \propto &\pi^{1/T_i}.
	    \e*
	    The higher the temperature is, the easier the exploration of the state space is for the associated chain.
	    Each iteration of the PT algorithm is decomposed into local and global moves. During local moves, each chain is updated independently of others. In particular, the $i$th chain is updated using a classical MCMC algorithm with stationary distribution $\pi_i$. For a global move, two chains $i$ and $j$ are randomly chosen and a swap of their current states is proposed, and accepted with the following Metropolis-Hastngs ratio:
	    \be*
	    \min\left\{1,\di\frac{\pi_i(x_j)\pi_j(x_i)}{\pi_i(x_i)\pi_j(x_j)}\right\},
	    \e*
	    where $x_i$ stands for the current state of the $i$th chain.

      \subsection{EES algorithm}
	    To use the EES algorithm introduced by \cite{KouZhouWong}, two sequences of $K$ temperatures and $K+1$ energy levels should be chosen: $H_1 < H_2 < \cdots < H_{K+1} =\infty$ and $T_1 =1 < T_2 < \cdots < T_K$, where $H_1 \leq \min(h(x))$ and $\pi(x) \propto \exp(-h(x))$.
	    A population of $K$ distributions with the following densities is considered:
	    \be*
	    \tpi_i(x) & \propto & \exp\{-h_i(x)\}, \qquad \textrm{where} \qquad h_i(x)  = \di\frac{\max\{h(x),H_i\}}{T_i}.
	    \e*
	    The main difference with the PT algorithm being the energy truncation. This energy truncation is used to flatten the distributions for easier exploration. This method uses energy rings for each chain, an energy ring containing past states of the chain of similar energy levels.
	    The algorithm begins by sampling the $K$th chain from a Metropolis-Hastings kernel with stationary distribution $\tpi_K$. Once convergence is reached, generated samples are stored in the energy rings of this $K$th chain, and the next chain targeting $\tpi_{K-1}$ starts. This $(K-1)$th chain will be updated by either (with probability $p_{ee}$) using a Metropolis-Hastings kernel with stationary distribution $\tpi_{K-1}$, or by proposing to replace the current state of the chain by a past state of the previous chain of similar energy level. This move corresponds to the equi-energy jump, and is based on the energy rings of the previous chain. Once convergence is reached, generated samples are stored in the energy rings of this $(K-1)$th chain, and the next chain targeting $\tpi_{K-2}$ starts. The EES algorithm successively steps down the energy and temperature ladder until the target distribution $\pi_1=\pi$ is reached. Each chain $i$, with $i<K$, is updated by either a Metropolis-Hastings kernel with stationary distribution $\tpi_i$ or by an equi-energy jump. More precisely, an equi-energy jump between two successive chains $i$ and $i-1$ is the following: a state $y$ is chosen from the chain $i$ such that $h(y)$ and $h(x_{i-1})$ belong to energy rings of similar energy. Then $y$ is accepted to be the next state of the ${(i-1)}$th chain with probability
	    \be*
	    \min\left\{1, \di\frac{\tpi_{i-1}(y)\tpi_i(x_{i-1})}{\tpi_{i-1}(x_{i-1})\tpi_i(y)}\right\}.
	    \e*

\section{PTEEM algorithm}
      \subsection{Description of the algorithm}
	    We introduce a sequence of $d+1$ energy levels $H_1 < H_2 < \cdots < H_{d+1} =\infty$ with $H_1 \leq \min(h(x))$, and a sequence of $N$ temperatures  $T_1=1 < T_2 < \cdots < T_N$.  The algorithm considers a population of $N$ chains associated with probability measures $\pi_i(x) \propto  \pi(x)^{1/T_i}$, each $\pi_i$ being a density with respect to a probability measure $\lambda$ on   $(\mathcal{X},\mathcal{B(X)})$, where $\mathcal X$ is a countably generated state space which coincides with the support of the $\pi_i$, and  $\mathcal{B(X)}$ stands for the associated Borel $\sigma$-algebra.

	    Energy rings $D_j, j=1,\ldots,d$ are constructed as follows: the state space ${\cal X}$ is partitioned according to the energy levels: ${\cal X}=\bigcup_{j=1}^d D_j$, where
	    \be*
	    D_j&=&\{x \in {\cal X}; h(x) \in [H_j,H_{j+1})\}, \ \  j= 2,\cdots, d \\ D_1&=&\{x \in {\cal X}; h(x) \in (-\infty,H_2)\}.
	    \e*
	    Compared to the energy rings of the EES method, these rings contain only current states, and there is only one sequence of energy rings for all the chains. By contrast the rings defined by \cite{KouZhouWong} contain past states, and a sequence of energy rings is constructed for each chain.

	    Each step of the PTEEM algorithm  is decomposed into two types of moves: local moves via classical MCMC algorithms and global moves allowing an exchange between two chains with similar energies.

	    \paragraph{Local moves}
	    Each chain is locally updated, independently of others. In particular, the $i$th chain is updated using one iteration of a classical MCMC algorithm with stationary distribution $\pi_i$. This algorithm could be a Metropolis-Hastings algorithm, a Gibbs sampler, an hybrid MCMC (\cite{MonteCarloStatMethods}), or a Reversible Jump MCMC (\cite{Green,RichardsonGreen}).
% 	    At each step a new state $y_i$ is proposed  to the  $i$th chain, for all $i=1,\cdots, N$, using MCMC algorithm.
% 	    When using  a Gibbs sampler,
% 	      the $i$th chain takes the value $y_i$.   When using Metropolis-Hasting algorithm,
% 	      the current value $x_i$ of the $i$th chain is replaced by $y_i$ with probability
% 	      \be*
% 	      \alpha&=&\min\left(1,\frac{f(y_i)q(y_i;x_i)}{f(x_i)q(x_i;y_i)}\right),
% 	      \e*
% 	      with $\pi(x)=f(x)/K$, where the normalizing constant $K$ may not be known. Assuming that the proposal distribution
% 	    is symmetric, i.e., $q(x; y) = q(y; x)$, the local move is accepted with probability  $\alpha=\min(1,\frac{f(y_i)}{f(x_i)})$.

	    \paragraph{Global moves}
	    At each step, an energy ring  $D_j$ containing at least two chains is chosen randomly. Two chains are then chosen uniformly  in $D_j$, say the $i$th and the $k$th ones (with $i<k$), and an exchange move is proposed between the current two states of these chains.
	    The move is from $s=(x_1,\cdots, x_i, \cdots,x_k,\cdots, x_N)$ to $s'=(x_1, \cdots, x_k, \cdots,x_i,\cdots, x_N)$.\\
	    The product $\sigma$-algebra is written $\cal B({\cal X})^N$, and the product measure is denoted by $\lambda_N$. The probability measure $\pi^*$ is defined as follows:
	    \begin{displaymath}
	    \pi^*(dx_1,dx_2,\ldots,dx_N)=\prod_{i=1}^{N} \pi_i(x_i)\lambda(dx_i) \qquad \textrm{on} \qquad ({\mathcal{X}}^N,\mathcal{B({\mathcal{X}})}^N)
	    \end{displaymath}
	    The probability acceptance for the global move is then given by:
	    \beg
	    \rho(s; s') & = &\min\left\{1,\di\frac{\pi^*(s')}{\pi^*(s)}\right\}
	    \nonumber
	    \\
	    & = &
	    \min\left\{1,\di\frac{\pi_i(x_k)\pi_k(x_i)}{\pi_i({x}_i)\pi_k(x_k)}\right\}.
	    \label{probaccep}
	    \en
	    Note that if the denominator is null, then the numerator is also null and by convention $\rho(s; s')$ is null.
	    The chains are not Markov by themselves, it is the whole stochastic process made of the $N$ chains together
	    that forms a Markov chain on $({\cal X}^N,{\cal B}({\cal X})^N)$.

	    \begin{rmk}
	    It is of interest to compare the total number of local and global moves required in PTEEM and EES algorithms.
	    Let us denote by $B$ the size of the burn-in period, by $R$ the number of iterations  necessary to initialize energy rings  within EES, and by $M$ the sample size required for the chain of interest (after the burn-in period).  We have:
	    \bit
	    \item
	    For EES, the total number of local moves is equal to
	    \be*
	    K(B+R)+(M-R) + (1-p_{ee})\left( \di\frac{(K-1)K}{2}(B+R)+(K-1)(M-R)\right),
	    \e*
	    and the total number of proposed global moves is equal to
	    \be*
	    p_{ee}\left( \di\frac{(K-1)K}{2}(B+R)+(K-1)(M-R)\right),
	    \e*
	    where $K$ denotes the number of chains in EES.
	    \item
	    For PTEEM, the total number of local moves is
	    $ NM + NB$, \\
	    and the total number of global moves is
	    $ M + B$, \\
	    where $N$ stands for the number of chains in PTEEM.
	    \eit
	    %In terms of number of moves, PTEEM necessitates more moves than EES when $M$ is large since usually $N$ is larger than $K$.
	    In terms of computational cost, we should take into account that in some problems the local algorithms used by EES and PTEEM can be different (see Section \ref{TFBS}), and thus can have different computational costs. In terms of storage, to obtain  the  $(i+1)$th iteration of the target chain, EES uses $KR+i + (1-p_{ee})\left( \di\frac{(K-1)KR}{2}+(K-1)i\right)$ values in memory to choose an element in an energy ring,  whereas PTEEM necessitates only $N$ values. Notice that CPU time to compute one iteration increases within EES as the simulations go along, while it is constant within PTEEM algorithm.
	    \end{rmk}

      \subsection{Some theoretical results}
	    In this section standard  sufficient conditions ensuring convergence of the PTEEM algorithm are given.
	    Denote by $S$ the Markov chain on $({\cal X}^N,{\cal B} ({\cal X})^N)$ obtained by the PTEEM algorithm, a state of $S$ is written $s$. The transition kernel associated with an iteration of PTEEM is written $P$, and  $P^k$ is the $k$-step transition kernel. They are defined on $({\cal X}^N \times {\cal B}({\cal X})^N)^2$. The transition kernel associated with the local move of the $i$th chain is written $PL_i$, and is defined on $(\mathcal{X} \times \mathcal{B}(\mathcal{X}))^2$. The transition kernel associated with the whole $N$ local moves of an iteration of PTEEM is written $PL$, and is defined on $(\mathcal{X}^N \times \mathcal{B}(\mathcal{X})^N)^2$. The transition kernel associated with the equi-energy move is written $PE$, and is defined on $({\cal X}^N \times {\cal B}({\cal X})^N)^2$. Writing
	    \begin{eqnarray*}
		    s &=& (x_1,\ldots, x_i,\ldots,x_k,\ldots,x_N)\\
		    s' &=& (x_1',\ldots, x_i',\ldots,x_k',\ldots,x_N'),
	    \end{eqnarray*}
	    we have
	    \begin{eqnarray*}
		    PL(s,s') &=& \prod_{i=1}^N PL_i(x_i,x_i'),\\
		    P(s,s') &=& (PE * PL)(s,s') = \int_{{\cal X}^N} PE(\tilde{s},s')PL(s,\tilde{s})d\tilde{s}.
	    \end{eqnarray*}
	    Write $q(s,s')$ the auxiliary distribution to propose $s'$ from $s$ in an equi-energy move, and $q_i(x_i,x_i')$ the auxiliary distribution to propose $x_i'$ from $x_i$ in a local move of the $i$th chain.
	    %Denote $\Theta$ the support of $\pi^*$. It can be written as $\Theta_1 \times \ldots \times \Theta_N$ with $\Theta_i \in \cal B(\cal X)$.\\
	    The total variation norm for a measure $\mu$ on $({\cal X}^N,\cal B({\cal X}^N)$ is defined by:
	    \begin{displaymath}
		    \| \mu\|_{TV}  =  \sup_{A \in \cal B ({\cal X}^N)} |\mu(A)|.%  - \inf_{A \in \cal B ({\cal X}^N)} \mu(A)
	    \end{displaymath}

	    \begin{prp} If the transition kernels associated with the local moves are reversible with stationary distributions $\pi_i$, $i=1,\cdots, N$, aperiodic and strongly $\lambda$-irreducible, then the chain ${S}$ is reversible, strongly $\lambda_N$-irreducible and we have for $\pi^*$-almost all $s \in {\cal X}^N$
	    \be*
	    \lim_{n\rightarrow \infty} \|P^n(s,.) - \pi^*\|_{TV} &=& 0.
	    \e*
	    Therefore $\pi^*$ is  the stationary distribution of $S$ and the chain associated with $T_1=1$  provides  samples corresponding to $\pi_1=\pi$, which is the target distribution.
	    \label{prop1}
	    \end{prp}
	    \textbf{Proof.} See Appendix \ref{proofprop1}.

	    \begin{rmk}
	    In Proposition \ref{prop1}, the transition kernels of the local moves are assumed to be aperiodic.
	    We can relax this hypothesis. In fact, it is sufficient that only one of the $N$ transition kernel is aperiodic to have
	    $P$  aperiodic.\\
	    The reversibility hypothesis can also be relaxed. If the reversibility of the local transition kernels is not assumed, the convergence results remained, but the reversibility of the chain $S$ is no more ensured. This reversibility can be interesting in order to use limit theorems (see \cite{MonteCarloStatMethods}).
	    \end{rmk}

	    This proposition has minimal assumptions, which are usually not difficult to verify, especially for classical MCMC algorithms like Metropolis-Hastings algorithms or Gibbs samplers.
	    However, it is possible to have a null set of states from which convergence does not occur.
	    The following lemma and proposition have stronger assumptions that ensure convergence from all starting points.

	    \begin{lmm}\label{lemme1}
	    Assume that the transition kernels associated with the local moves are reversible with stationary distributions $\pi_i$, $i=1,\cdots, N$, aperiodic and strongly $\lambda$-irreducible, and assume the strict positivity of the density $\pi^*$ on $\mathcal{X}^N$ ($\forall s \in \mathcal{X}^N, \pi^*(s)>0$). Then the chain ${S}$ is reversible, strongly $\lambda_N$-irreducible, positive and Harris-recurrent.
	    \end{lmm}
	    \textbf{Proof.} See Appendix \ref{prooflemme1}.\\

	    \noindent The following proposition is a consequence of Lemma \ref{lemme1}.

	    \begin{prp} Assume that the transition kernels associated with the local moves are reversible with stationary distributions $\pi_i$, $i=1,\cdots, N$, aperiodic and strongly $\lambda$-irreducible, and assume the strict positivity of the density $\pi^*$ on $\mathcal{X}^N$ ($\forall s \in \mathcal{X}^N, \pi^*(s)>0$). Then we have for all $s \in {\cal X}^N$
	    \be*
	    \lim_{n\rightarrow \infty} \|P^n(s,.) - \pi^*\|_{TV} &=& 0.
	    \e*
	    \label{prop2}
	    \end{prp}
	    \textbf{Proof}: Using Lemma \ref{lemme1} and Proposition \ref{prop1}, $S$ is a Markov chain $\pi^*$-irreducible, aperiodic, with stationnary distribution $\pi^*$ and Harris-recurrent. The result follows from Theorem 1 of \cite{Tierney94}.
	    \cqfd \\

%	    Obviously, the remarks previously done on the assumptions of \ref{prop1}  can be applied on the assumptions of \ref{lemme1} and \ref{prop2}.
%	    Notice that under assumptions of \ref{lemme1} and \ref{prop2}, the ergodic theorem which is an equivalent of the law of large numbers for Markov chains can be applied.

      \subsection{Calibration}\label{ChoiceTempEnergy}
      Ideally, the more chains and energy rings are used, the better the result of the algorithm will be. However, it is not always possible in practice for computational reasons. That is why we suggest from our experience simple ways to choose the number of chains and energy rings, and to calibrate the energy ladder and the temperatures.

	      \paragraph{Number of rings}
		    The number of energy rings $d$ should be chosen in relation with the complexity of the target density. For instance, if the amplitude between the larger and the lower energy levels is large, or if the different modes are associated with different energy levels, then it is necessary to increase the number of rings.

	      \paragraph{Energy ladder}
		    The levels $H_1,H_2,\ldots,H_d$ determine the energy rings. The first energy ring includes states having an energy level lower than $H_2$, and ideally only few states having an energy level lower than $H_1$. The last energy ring includes states having an energy value higher than $H_d$.
		    To choose $H_1$ and $H_d$ we use one or few runs of a classical MCMC algorithm  with target density $\pi$.
		    We take for $H_d$ the energy associated with a state with high enough finite energy compared to other states. Concerning  $H_1$, we
		    take the energy corresponding to an observed mode. In practice, we can take for $H_d$ the energy associated with a state after few iterations of the algorithm, and for $H_1$ the energy associated with a state after a burn-in period.\\
		    Once the values $H_1$ and $H_d$ are chosen, the $\ln(H_i)$ or the $\ln(H_{i+1}-H_i)$ can be set to be evenly spaced.% on a logarithmic scale
% 		      \begin{displaymath}
% 			      ln(H_i) = ln(H_1) + i \frac{ln(H_d)-ln(H_1)}{d-1}.
% 		      \end{displaymath}
		    \begin{rmk}
		     Concerning  $H_1$, if the modes of the distribution of interest are known, we just have to take $H_1$ slightly lower than the energy of the highest mode.
		    \end{rmk}
		
	      \paragraph{Number of chains}
		     If the number of chains $N$ is chosen too small, the chance of having samples from different chains in the same ring will be too small. From our experience, $N$ should be at least equal to 3$d$, and choosing it between 3$d$ and 5$d$ is usually satisfactory. However, one can always use more chains if the computation time is not a problem.
		      	
	      \paragraph{Temperatures}
		      The distribution associated with the highest temperature should be sufficiently flattened so that the associated chain can move freely from one mode to another. After choosing a $T_{N}$ value we just have to check that the associated chain moves easily. %(with a run of a classical MCMC algorithm with target density $\pi$)
		      $T_{1}$ is obviously equal to 1, and is associated with the chain of interest. Once $T_{1}$ and $T_{N}$ are fixed, the other temperatures can be chosen by evenly spacing them on a logarithmic scale, by evenly spacing their inverses, or by evenly spacing their inverses geometrically (see for instance \cite{KouZhouWong}, \cite{NagataWatanabe} or \cite{Neal1996}). Following \cite{Atchade2010}, we can try to adjust the temperatures so that the proportion of accepted equi-energy moves is approximately 0.234.\\
		      We tried simple ways to choose the energy levels in combination with the temperatures, but none of them gave conclusive results. However, if the expression of the target density is known, it could be possible to choose theoretically the temperatures and the energy levels so that each ring contains in mean the same number of chains.
		      % by computing the probability that a chain at fixed temperature $T$ belongs to a given ring.

	      \paragraph{Checking that the choices of temperatures and energy ladder are relevant}
		      It is necessary to check on a run of PTEEM that the choices of temperatures and energy ladder are relevant. The chain 1 should have almost all its states in the first energy ring, and the last chain should have almost all its states in the last energy ring. For the other chains, the distribution in the rings can be considered as correct if there is no "energy gap" between adjacent chains, and if each chain performed equi-energy moves with chains having higher and lower temperatures. If this is not the case, poor mixing is observed between chains, and it is then necessary to adjust the temperatures or the energy levels, adding new temperatures for instance or proposing a new calibration. This problem of calibration is illustrated in Table \ref{tab1}.% and in an example of Section \ref{4.1}.\ref{•}
		
			      \begin{table}[!h]
			      \begin{center}
			      \begin{tabular}{|c|c|c|c|c|c||c|c|c|c|c|}
			      \hline
			      & \multicolumn{5}{|c||}{Bad repartition} & \multicolumn{5}{|c|}{Good repartition} \\
			      \hline
			      Energy ring & 1 & 2 & 3 & 4 & 5 &   1 & 2 & 3 & 4 & 5 \\
			      \hline
			      chain $i-2$ & 990 & 10 & 0 & 0 & 0  & 990 & 10 & 0 & 0 & 0 \\
			      chain $i-1$ & 950 & 50 & 0 & 0 & 0  & 701 & 202 & 97 & 0 & 0\\
			      chain $i$ & 900 & 100 & 0 & 0 & 0  & 387 & 408 & 205 & 0 & 0 \\
			      chain $i+1$ & 0 & 2 & 237 & 511 & 250 & 45 & 312 & 355 & 288 & 0\\
			      chain $i+2$ & 0 & 0 & 105 & 610 & 285  & 0 & 64 & 517 & 353 & 66 \\
			      \hline
			      \end{tabular}
			      \caption{Illustration for  bad and good repartitions of the states in the energy rings. There is an energy gap between chains $i$ and $i+1$ in the bad repartition case.}
			      \label{tab1}
			      \end{center}
			      \end{table}

\section{Example of simulations using local Metropolis-Hastings moves}\label{SimuMH}
	To compare the three algorithms (PT, EES and PTEEM) when the local move is a Metropolis-Hastings algorithm, we consider sampling from a two-dimensional normal mixture model taken from \cite{LiangWong} and used as an illustration by \cite{KouZhouWong}. Let
	\begin{displaymath}
	f(x) =\sum_{i=1}^{20} \frac{w_i}{\sigma_i \sqrt{2\pi}} exp\Big(-\frac{1}{2\sigma_i^2}(x-\mu_i)'(x-\mu_i)\Big),
	\end{displaymath}
	where $\sigma_1=\ldots=\sigma_{20}=0.1$, $w_1=\ldots=w_{20}=0.05$, and the 20 mean vectors
	\begin{eqnarray*}
	(\mu_1,\ldots,\mu_{20}) &=& \left(
		\begin{array}{c c c c c c c c c c}
			2.18 & 8.67 & 4.24 & 8.41 & 3.93 & 3.25 & 1.70 & 4.59 & 6.91 & 6.87\\
			5.76 & 9.59 & 8.48 & 1.68 & 8.82 & 3.47 & 0.50 & 5.60 & 5.81 & 5.40
		\end{array} \right.\\
		 & & \left.
		\begin{array}{c c c c c c c c c c}
			5.41 & 2.70 & 4.98 & 1.14 & 8.33 & 4.93 & 1.83 & 2.26 & 5.54 & 1.69\\
			2.65 & 7.88 & 3.70 & 2.39 & 9.50 & 1.50 & 0.09 & 0.31 & 6.86 & 8.11
		\end{array} \right).
	\end{eqnarray*}
	The different local modes are quite far from each other (most of them are more than 15 standard deviations from the nearest ones), hence this mixture distribution is quite challenging for sampling. In addition, the initial states of the different chains were drawn from a uniform distribution on $[0,1]^2$, a region far from the local modes.\\
	Each algorithm was run 100 times. For each run, the PT and PTEEM algorithms were run for 2500 iterations after a burn-in period of 2500 iterations. Similarly, for each chain of the EES the burn-in period was of 2500 iterations, and for the first chain (the target chain) 2500 iterations were simulated after this burn-in period and the period to construct the rings, which was of 500 iterations. As in \cite{KouZhouWong}, the Metropolis-Hastings proposal was  a bivariate Gaussian $X_{n+1}^{(i)} \sim \mathcal{N}_2(X_n^{(i)},\tau_i^2 I_2)$, with $\tau_i=0.25 \sqrt{T_i}$. Unlike them, the step size $\tau_i$ was not tuned later in the algorithms such that the acceptance ratio is in the range (0.22,0.32). Indeed, we would like to compare algorithms as simple as possible.
	For the EES, we took the same number of chains, the same energy levels, the same temperatures and the same equi-energy jump probability than \cite{KouZhouWong} ($K=5$, $H=(0.2,2,6.3, 20, 63.2)$, $T=(1,2.8,7.7,21.6,60)$, $p_{ee}=0.1$). For the PT and PTEEM algorithms,  $N=20$ chains were taken, with temperatures between 1 and 60 evenly spaced on a logarithmic scale. As in \cite{KouZhouWong}, the PT algorithm used a swap between neighboring temperature chains for the exchange operation, but only one swap was proposed at each iteration, to make it comparable with the PTEEM. For the PTEEM, the same 5 groups of energy than for the EES were taken.

	Mean acceptance rates for the local Metropolis-Hastings moves and for the exchange moves between chains for the three algorithms are given in Table \ref{tabTxAccept}. In comparison \cite{KouZhouWong} obtained results slightly different probably because the step size $\tau_i$ was tuned in their EES.\\
	\begin{table}[!h]
	\begin{center}
	\begin{tabular}{|c|c|c|}
	\hline
	 & Local moves & Exchange moves \\
	\hline
	EES & 0.387 & 0.799\\
	PT & 0.337 & 0.905 \\
	PTEEM & 0.333 & 0.822\\
	\hline
	\end{tabular}
	\caption{Mean acceptance rates for local moves  and exchange moves on 100 runs, for EES, PT and PTEEM algorithms.
	}\label{tabTxAccept}
	\end{center}
	\end{table}
	\indent To compare the ability of each algorithm to explore the distribution space, we considered for each run of each algorithm the number and frequency of visited modes by the target chain, as well as the estimations of the mean vector $(E(X_1),E(X_2))$ and of the second moments $(E(X_1^2),E(X_2^2))$ using the samples generated from the target chain. Table \ref{tabMoments} (A) contains these estimations. Concerning the estimations of the mean vector and of the second moments, the EES and  PTEEM estimates were more accurate than those of the PT, with smaller mean squared errors. Moreover, it appeared that the PTEEM estimates were slightly more accurate than those of the EES.
	\begin{table}[!h]
	\begin{center}
	\begin{tabular}{|c|c|c|c|c|c|}
	\hline
	\multicolumn{2}{|c|}{} & $E(X_1)$ & $E(X_2)$ & $E(X_1)^2$ & $E(X_2)^2$\\
	\multicolumn{2}{|c|}{True value} & 4.478 & 4.905 & 25.605 & 33.920\\
	\hline	
	& EES & 4.448 (0.301) & 4.953 (0.458) & 25.229 (3.112) & 34.226 (4.507)\\
	& PT & 3.971 (0.809) & 4.137 (1.114) & 21.510 (7.741) & 27.510 (10.407)\\
	\multirow{-3}{*}{(A)} & PTEEM & 4.483 (0.324) & 4.912 (0.454) & 25.556 (3.366) & 33.889 (4.406)\\
	\hline
	& EES & 5.088 (0.373) & 6.001 (0.515) & 32.005 (4.086) & 45.306 (5.638)\\
	\multirow{-2}{*}{(B)} & PTEEM & 4.745 (0.365) & 5.468 (0.491) & 28.908 (3.774) & 40.617 (5.019)\\
	\hline
	\end{tabular}
	\caption{Estimations of the mean vector $(E(X_1),E(X_2))$ and of the second moments $(E(X_1^2),E(X_2^2))$ using the samples generated from the target chain, obtained on 100 runs for  EES, PT and PTEEM algorithms. The standard deviations are given between parentheses. (A) corresponds to the case with equal variances, and (B) to the case with unequal variances.}\label{tabMoments}
	\end{center}
	\end{table}
	Concerning the number of visited modes, good results were obtained by the EES and PTEEM algorithms compared to the PT. The results are reported in Table \ref{tabModesVisites}. The mean number of visited modes by the PT on the 100 runs was  14.31, compared to 19.92 for the EES and 19.98 for the PTEEM.
	\begin{table}[!h]
	\begin{center}
	\begin{tabular}{|c|c|c|}
	\hline
	PT & EES & PTEEM\\
	\hline
	2 to 10 missed. & 1 missed for 4 runs. &  1 missed for 2 runs.\\
	A mean of 5.69 missed. & 2 missed for 2 runs. &  \\
	\hline
	\end{tabular}
	\caption{Number of missed modes by the 100 runs for EES, PT and PTEEM algorithms.}\label{tabModesVisites}
	\end{center}
	\end{table}
	Then, as in \cite{KouZhouWong}, we counted in each of the 100 runs for the three algorithms how many times the target chain visited each mode in the last 2500 iterations. The absolute frequency error is given by  $err_i=|\hat{f_i}-0.05|$, where $\hat{f_i}$ is the sample frequency of the $i${th} mode being visited ($i=1,\ldots,20$). The median and the maximum of $err_i$ over the 100 runs was calculated. To compare the three algorithms the ratios of these values between PT and EES, between PT and PTEEM and between EES and PTEEM were calculated for each mode. All these ratios are presented in Table \ref{tabFrequences}.
	As denoted in \cite{KouZhouWong},  EES seemed to be more efficient than  PT: the mean of the ratios $R_{med( PT/EES)}$ over the 20 modes was 2.42, and the mean of the ratios $R_{max(PT/EES)}$ over the 20 modes was 2.92. As expected, PTEEM gave better results than PT: the mean of  $R_{med(PT/PTEEM)}$ was 2.52, and the mean of  $R_{max(PT/PTEEM)}$  was 3.07. Besides, we noticed a slight improvement of  PTEEM compared to EES: 1.05 for the  mean of $R_{med(EES/PTEEM)}$, and 1.13 for the mean of $R_{max(EES/PTEEM)}$.
	\begin{table}[!h]
	\begin{small}
	\begin{center}
	\begin{tabular}{c c c c c c c c c c c}
	\hline
	 & \textbf{$\mu_1$} &  \textbf{$\mu_2$} & \textbf{$\mu_3$} & \textbf{$\mu_4$} & \textbf{$\mu_5$} & \textbf{$\mu_6$} & \textbf{$\mu_7$} & \textbf{$\mu_8$} & \textbf{$\mu_9$} & \textbf{$\mu_{10}$}\\
	\hline
  	$R_{med}$ PT/EES & 2.16 & 2.80 & 2.92 & 2.22 & 1.98 & 2.21 & 3.10 & 2.07 & 2.07 & 2.69 \\
  	$R_{max}$ PT/EES & 3.59 & 2.61 & 2.81 & 2.10 & 1.55 & 2.43 & 2.54 & 1.53 & 2.93 & 4.50 \\
  	$R_{med}$ PT/PTEEM & 2.60 & 3.72 & 2.63 & 2.19 & 1.79 & 2.97 & 2.77 & 2.55 & 2.32 & 2.64 \\
  	$R_{max}$ PT/PTEEM & 3.44 & 1.76 & 2.27 & 2.30 & 2.44 & 3.06 & 5.23 & 2.92 & 2.83 & 5.23 \\
  	$R_{med}$ EES/PTEEM & 1.21 & 1.33 & 0.90 & 0.99 & 0.91 & 1.35 & 0.89 & 1.23 & 1.12 & 0.98 \\
  	$R_{max}$ EES/PTEEM & 0.96 & 0.67 & 0.81 & 1.09 & 1.58 & 1.26 & 2.06 & 1.91 & 0.96 & 1.16 \\
	\hline
	& \textbf{$\mu_{11}$} &  \textbf{$\mu_{12}$} & \textbf{$\mu_{13}$} & \textbf{$\mu_{14}$} & \textbf{$\mu_{15}$} & \textbf{$\mu_{16}$} & \textbf{$\mu_{17}$} & \textbf{$\mu_{18}$} & \textbf{$\mu_{19}$} & \textbf{$\mu_{20}$} \\
	\hline
  	$R_{med}$ PT/EES & 2.51 & 2.46 & 2.77 & 2.63 & 2.39 & 1.76 & 3.06 & 2.22 & 2.10 & 2.37 \\
  	$R_{max}$ PT/EES & 4.58 & 1.60 & 3.23 & 4.61 & 3.26 & 2.10 & 2.83 & 4.77 & 3.50 & 1.36 \\
  	$R_{med}$ PT/PTEEM & 2.14 & 1.98 & 1.79 & 2.84 & 2.75 & 2.18 & 2.72 & 2.78 & 2.43 & 2.60 \\
  	$R_{max}$ PT/PTEEM & 3.05 & 2.02 & 2.35 & 4.16 & 3.44 & 1.79 & 3.72 & 3.78 & 3.50 & 2.16 \\
  	$R_{med}$ EES/PTEEM & 0.85 & 0.81 & 0.65 & 1.08 & 1.15 & 1.24 & 0.89 & 1.25 & 1.16 & 1.10 \\
  	$R_{max}$ EES/PTEEM & 0.67 & 1.26 & 0.73 & 0.90 & 1.06 & 0.85 & 1.32 & 0.79 & 1.00 & 1.58 \\
	\hline
	\end{tabular}
	\end{center}
	\caption{For each mode, ratios of median ($R_{med}$) and ratios of maximum ($R_{max}$) are  for PT over  EES,  PT over  PTEEM, and  EES over PTEEM. Each ratio is obtained on 100 runs.}\label{tabFrequences}
	\end{small}
	\end{table}

\begin{figure}[!h]
	\begin{center}
		\includegraphics[scale=0.6]{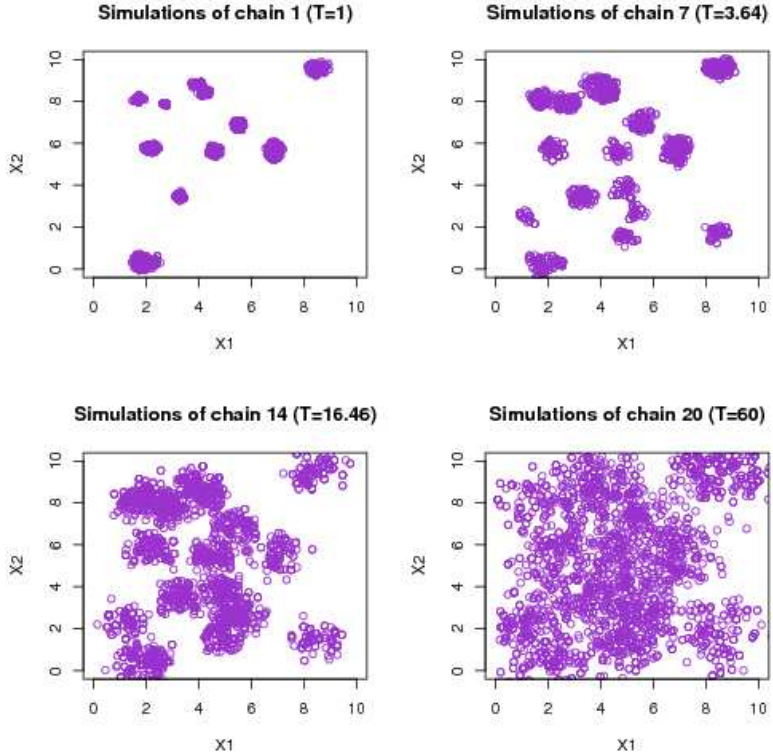}
		\caption{Simulations for chains 1, 7, 14 and 20 obtained by one run of the PT algorithm.}\label{fig:MCMCMC1}
	\end{center}
	\end{figure}

\begin{figure}[!h]
	\begin{center}
		\includegraphics[scale=0.6]{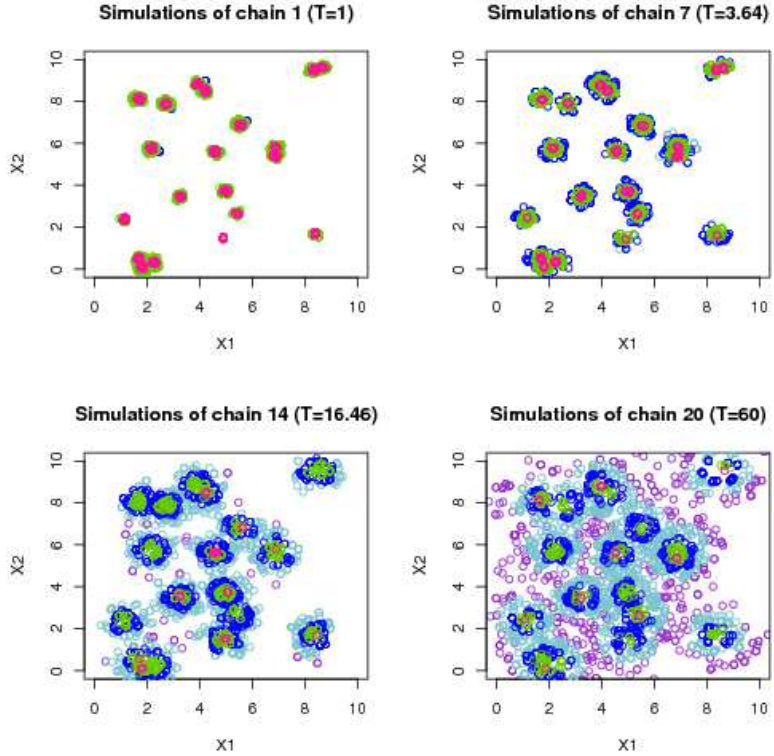}
		\caption{Simulations for chains 1, 7, 14 and 20 obtained by one run of the PTEEM algorithm. The colors correspond to the five energy levels.}\label{fig:PTEEM1}
	\end{center}
	\end{figure}

\begin{figure}[!h]
	\begin{center}
		\includegraphics[scale=0.6]{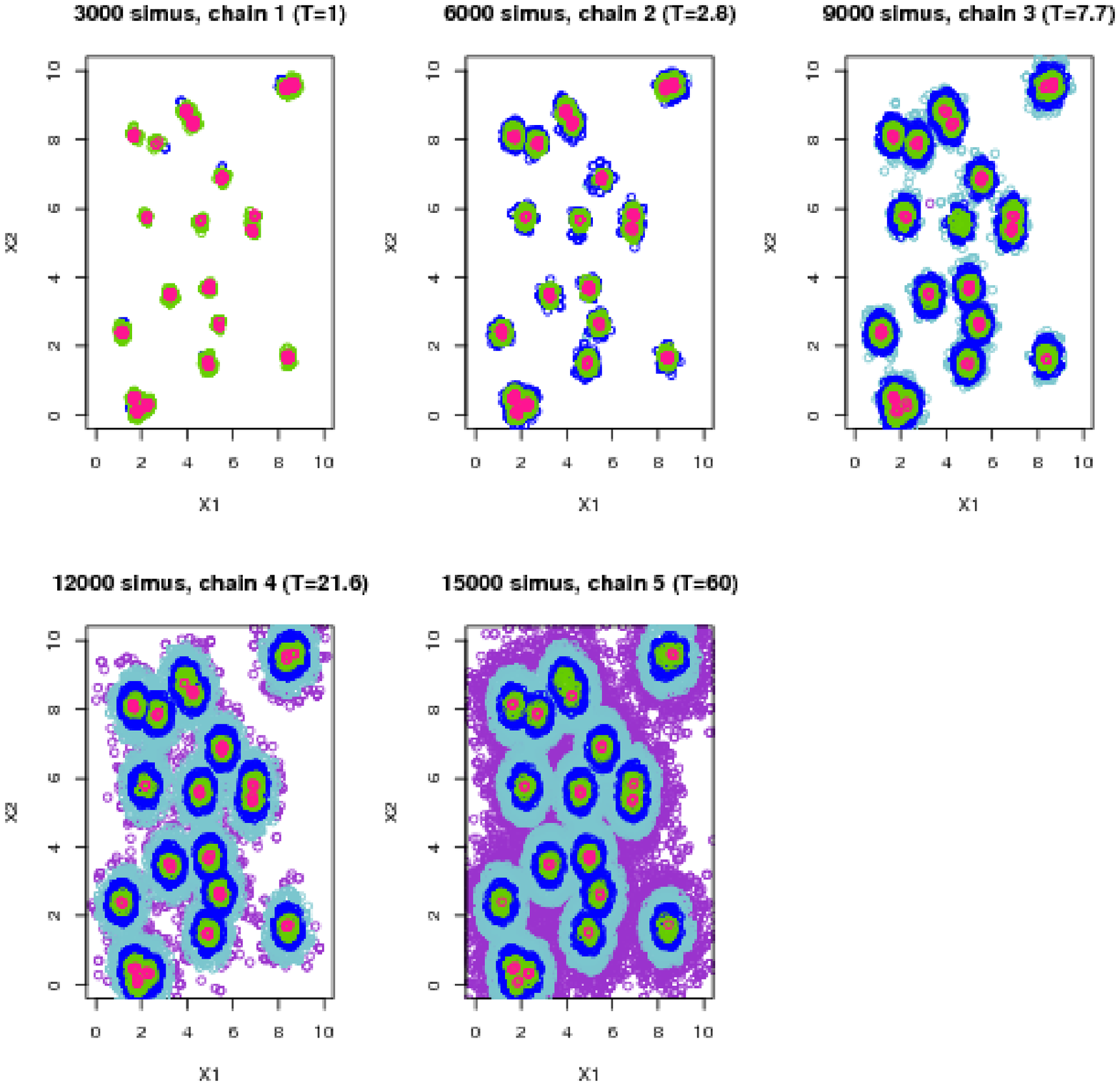}
		\caption{Simulations for chains 1 to 5 obtained by one run of the EES. The colors correspond to the five energy levels.}\label{fig:EES1}
	\end{center}
	\end{figure}

 	Figures \ref{fig:MCMCMC1} and \ref{fig:PTEEM1} show the last 2500 iterations after burn-in for the chains 1, 7, 14 and 20 obtained by one run of the PT algorithm, and by one run of the PTEEM algorithm. Figure \ref{fig:EES1} shows the simulations after a burn-in period for chains 1 to 5 obtained by a run of EES.
	The first chains  of the PTEEM and EES visited all the modes of the target density whereas the first chain of PT did not visit all of them. Notice that chains with the highest temperatures of the PT algorithm visited all the modes, and these chains for the EES  kept in memory lots of iterations.

	Table \ref{tabEEmoves} presents  the repartition of accepted equi-energy moves for chains 1, 10 and 20, with other possible chains within a run of the PTEEM algorithm.
	As expected, the closer the temperatures of chains were, the more often the  equi-energy moves were accepted.
	Note that equi-energy moves had been proposed and accepted for all possible pairs of chains, including for pairs of chains with very different  temperatures.

	\begin{table}[!h]
	\begin{center}
	\begin{tabular}{|c|c|c|c|}
	\hline
		& chain 1 & chain 10 & chain 20\\
	\hline
	chain 1 & 0.00 & 4.63 & 0.50 \\
	chain 2 & 16.32 & 4.33 & 0.62 \\
	chain 3 & 14.34 & 4.29 & 0.64 \\
	chain 4 & 11.98 & 4.64 & 0.70 \\
	chain 5 & 9.96 & 4.89 & 0.76 \\
	chain 6 & 8.26 & 5.46 & 1.03 \\
	chain 7 & 6.57 & 5.76 & 1.17 \\
	chain 8 & 6.01 & 6.26 & 1.50 \\
	chain 9 & 4.96 & 6.74 & 2.03 \\
	chain 10 & 4.32 & 0.00 & 2.30 \\
	chain 11 & 3.25 & 7.11 & 3.13 \\
	chain 12 & 2.85 & 6.67 & 4.33 \\
	chain 13 & 2.42 & 6.65 & 5.61 \\
	chain 14 & 1.98 & 6.11 & 7.38 \\
	chain 15 & 1.61 & 5.76 & 8.75 \\
	chain 16 & 1.44 & 5.13 & 10.64 \\
	chain 17 & 1.15 & 4.64 & 13.72 \\
	chain 18 & 1.08 & 4.32 & 16.09 \\
	chain 19 & 0.87 & 3.63 & 19.10 \\
	chain 20 & 0.62 & 2.99 & 0.00 \\
	\hline
	\end{tabular}
	\caption{Repartition (in \%) of accepted equi-energy moves between chain 1 and other possible chains (mean on 100 runs of PTEEM). Idem for chains 10 and 20.}\label{tabEEmoves}
	\end{center}
	\end{table}

	As in \cite{KouZhouWong}, it appeared that the EES algorithm gave better results than the classical PT. Besides the PTEEM algorithm gave results comparable to those of the EES, and even slightly better. In this example all the modes have exactly the same energy and the same component variance.

	\paragraph{Case of unequal variances and energy levels}~~\\
	In order to study the behavior of the PTEEM and the EES algorithms in case of unequal variances and energy levels, we took $\sigma_1=\ldots=\sigma_5=0.4$, $\sigma_6=\ldots=\sigma_{13}=0.1$ and $\sigma_{14}=\ldots=\sigma_{20}=0.05$. The modes associated with small variances have lower energy levels, and those associated with large variances have higher energy levels.\\
	The PTEEM and EES algorithm was run 100 times. For each run, the PT and PTEEM algorithms were run with the same number of iterations and the same Metropolis-Hastings proposal as previously. For the PTEEM, we still used 20 chains, but with six energy levels: we took $H_1=0.5$ and $H_2$ to $H_6$ evenly spaced on a logarithmic scale between 1.5 and 20. The temperatures were still evenly spaced on a logarithmic scale between 1 and 60. For the EES, six chains were used with the same energy levels than for the PTEEM, the temperatures were also evenly spaced on a logarithmic scale between 1 and 60, and $p_{ee}=0.1$. It is interesting, because the 20 modes were divided into three energy rings, and the second energy ring contained modes with different variances ($0.4^2$ and $0.1^2$). \\
	Concerning the estimations of the mean vector $(E(X_1),E(X_2))$ and of the second moments $(E(X_1^2),E(X_2^2))$, Table \ref{tabMoments} (B) shows that the PTEEM estimates were more accurate than those of the EES, with smaller mean squared errors. The mean number of visited modes by the PTEEM on the 100 runs was 18.91, compared to 17.83 for the EES. Concerning the absolute frequency errors, we noticed an improvement of PTEEM compared to EES, as the mean of the ratios $R_{med(EES/PTEEM)}$ over the 20 modes was 1.213, and the mean of the ratios $R_{max(EES/PTEEM)}$ was 1.203.\\
	Note that if some components are associated to very small variances ($\sigma=0.01$ for instance), it became difficult for both the PTEEM and the EES algorithms to detect these modes when they are isolated, far from other ``larger'' modes.

	\begin{rmk}
	   In this last example, the EES algorithm necessitates 72 250 local moves and 5750 global moves in mean, while the PTEEM algorithm necessitates 100 000 local moves and 5000 global moves. The goal was to study these two algorithms when they use the same number of energy rings (6 here) and with the same number of iterations after burn-in and rings construction for the chain of interest. Using 20 chains for the PTEEM matches the advices given in section \ref{ChoiceTempEnergy}. %Computationnally we did not notice a big difference between a run of the EES and a run of the PTEEM, as the EES took more time for higher order iterations because of all samples kept in memory.
	\end{rmk}

\section{Example of estimation using local Gibbs samplers moves}\label{Galaxy}
	  We consider estimation of model parameters in case of a mixture model with known number of components. The classical algorithm used for this kind of problem is a Gibbs sampler. However, some difficulties are encountered to combine  the original EES with a Gibbs sampler. Therefore, we compared  only performances of PT and PTEEM algorithms, using the well-known example of the Galaxy dataset (see for instance \cite{RichardsonGreen}).

%	  Two illustrations of Gaussian mixtures are treated: an example with simulated data, and the well-known example of the Galaxy dataset.\\
	  \noindent We consider independent observations $y_1, \cdots, y_n$ from $k$ mixture components
	  \begin{displaymath}
	  y_i \sim \sum_{j=1}^k w_j f(.|\mu_j,\sigma_j^2), \qquad i=1,\ldots,n,
	  \end{displaymath}
	  with $k$ fixed and known and where $f(.|\mu_j,\sigma_j^2)$ denotes the density of the Gaussian distribution $\mathcal{N}(\mu_j,\sigma^2_j)$. The sizes of the $k$ groups are proportional to  $w_1,w_2,\ldots,w_k$, which are the weights of the components. The parameters to be estimated are the means $\mu_j$, the variances $\sigma^2_j$, and the weights $w_j$, for $ j=1,\ldots,k$.
	  The label of the component from which each observation is drawn is unknown, and  a label vector $c$ which is a latent allocation vector is introduced as follows: $c_i=j$ if the observation $y_i$ is drawn from the $j$th component. The variables $c_i$ are supposed independent with distributions
	  \begin{displaymath}
		  p(c_i=j)=w_j, \qquad j=1,\ldots,k.
	  \end{displaymath}
	  Write $y=(y_i)_{i=1, \ldots, n}$, $\mu=(\mu_j)_{j=1, \ldots, k}$, $\sigma^2=(\sigma^2_j)_{j=1, \ldots, k}$, $w=(w_j)_{j=1, \ldots, k}$ and $c=(c_i)_{i=1, \ldots, n}$. The $\mu_j$ and $\sigma_j^{-2}$ are supposed to be independent  with the following priors:
	  \begin{equation}\label{prior}
		  \mu_j \sim \mathcal{N}(\xi,\kappa^{-1}),  \qquad \sigma_j^{-2} \sim \Gamma(\alpha,\beta) \qquad \textrm{and} \qquad \beta \sim \Gamma(g,h),
	  \end{equation}
	  where $\beta$ and $h$ are rate parameters. The prior on $w$ is taken as a symmetric Dirichlet distribution
	  \begin{displaymath}
		  w \sim D(\delta,\delta,\ldots,\delta).
	  \end{displaymath}
	  The parameters $\delta$, $\xi$, $\kappa$, $\alpha$, $g$ and $h$ are supposed to be fixed. Let us denote by  $m_j=\sum_{i=1}^n \mathds{1}_{c_i=j}$ the number of observations labeled by $j$.
	  The joint posterior density, the full conditional distributions and the formula of the acceptance rate for the equi-energy move are given in Appendix \ref{formulesex}.\\
	  In this example, the estimates of the parameters obtained after labeling were quite good and similar for the PT and PTEEM algorithms. They were even comparable to those obtained with a classical Gibbs sampler. The major difference between these three algorithms was the ability to explore the parameter space: the Gibbs sampler stayed in local modes for many successive iterations, while the PT and PTEEM algorithms easily jumped from one mode to another. Consequently, we focused on the label-switching phenomenon (see \cite{JasraStephensHolmes2005}), and not on the estimation of the parameters.\\

		The data  consist of the velocities of 82 distant galaxies diverging from our own. We fix the number of components to $k=6$, and we took for the fixed parameters in (\ref{prior}): $\alpha=3$, $\xi=20$, $\delta=1$, $\kappa=1/R^2$, $g=0.2$ and $h=10/R^2$, where $R=10$. The algorithms PT and PTEEM were run 100 times, each run consisting of 10000 iterations after a burn-in period of 2000 iterations. We used $20$ chains and $5$ energy rings. As in the previous example, the PT algorithm used a swap between neighboring temperature chains for the exchange operation, and only one swap was proposed at each iteration.
		Concerning the energy ladder, after a run of a classical Gibbs sampler with  target density $\pi$, we
		chose $H_1=180 $ and $H_5=260$. Four energy rings were obtained with levels evenly spaced between $H_1$  and $H_5$ on a logarithmic scale, the fifth ring containing all states having an energy value higher than $H_5$. The levels obtained were 180, 197.3, 216.3, 237.2 and 260.
		We chose $N=20$ temperatures between 1 and 4, with their inverses evenly spaced. %We get 1.00, 1.04, 1.09, 1.13, 1.19, 1.25, 1.31, 1.38, 1.46, 1.55, 1.65, 1.77, 1.90, 2.05, 2.24, 2.45, 2.71, 3.04, 3.45 and 4.00.\\
		Table \ref{tabPTEEMRings2} shows for several chains the distributions of states in the energy rings.\\
		\begin{table}[!h]
		\begin{center}
		\begin{tabular}{|c|c|c|c|c|c|}
		\hline
		 & $(-\infty,197.3)$ & $[197.3,216.3)$ & $[216.3,237.2)$ & $[237.2,260)$ & $[260,+\infty)$\\
		\hline
		chain 1 & 9602 & 396 & 2 & 0 & 0 \\
		chain 4 & 4487 & 5343 & 170 & 0 & 0 \\
 		chain 8 & 225 & 6123 & 2863 & 768 & 21\\
		chain 10 & 5 & 990 & 3528 & 5017 & 460\\
		chain 12 & 0 & 50 & 1047 & 6662 & 2241\\
		chain 16 & 0 & 0 & 5 & 2266 & 7729 \\
		chain 20 & 0 & 0 & 0 & 312 & 9688 \\
		\hline
		\end{tabular}
		\caption{Distribution in the energy rings of states from 10000 iterations, for one run of PTEEM and for chains 1, 4, 8, 10 12, 16 and 20.}\label{tabPTEEMRings2}
		\end{center}
		\end{table}
		Clearly, the mixture posterior has $k!=720$ symmetric modes and, in theory, for a very high number of iterations, the chain of interest should have visited all modes, with equal frequencies. When the chain goes from one mode to another, there is the so-called label-switching phenomenon (see \cite{JasraStephensHolmes2005}). Such a   phenomenon is a useful convergence diagnostic to check if the chain of interest has explored all possible labelings of the parameters. To compare PT and PTEEM algorithms we considered for each run of each algorithm both the number and the frequency of visited modes by the target chain. Table \ref{tabPTEEMgalaxy} shows that on 100 runs of PTEEM the target chain visited more modes than on 100 runs of PT. Hence the label-switching phenomenon seems to occur more often during a run of PTEEM than during a run of PT.
		\begin{table}[!h]
		\begin{center}
		\begin{tabular}{|c|c|c|c|c|}
		\hline
		 & mean & standard deviation & min & max\\
		\hline
		PT & 645.04 & 13.52 & 610 & 683 \\
		PTEEM & 666.52 & 9.23 & 641 & 692\\
		\hline
		\end{tabular}
		\caption{Means, standard deviations, minimal and maximal values of the number of visited modes, on 100 runs of PT and PTEEM.}.\label{tabPTEEMgalaxy}
		\end{center}
		\end{table}
		We also counted in each of the 100 runs for the two algorithms how many times the target chain visited each mode in the last 10000 iterations. The absolute frequency error is given by $err_i=|\hat{f_i}-1/6!|$, where $\hat{f_i}$ is the sample frequency of the $i${th} mode being visited ($i=1,\ldots,6!$). We then calculated the mean and median of this absolute frequency error over the 100 runs and the 6! modes.
		Absolute frequency errors were slightly lower for PTEEM with a mean (resp. a median) of 0.119\% (resp. 0.099\%), compared to 0.126\% (resp. 0.099\%) for PT.

		We studied further the equi-energy moves of the algorithm {PTEEM}. In Table \ref{tabEEmoves3} it appears that  exchange moves were more frequent between chains with similar temperatures.
		\begin{table}[!h]
		\begin{center}
		\begin{tabular}{|c|c|c|c|}
		\hline
		 & chain 1 & chain 10 & chain 20\\
		\hline
		chain 1 & 0.00 & 0.02 & 0.00 \\
		chain 2 & 63.65 & 0.10 & 0.00 \\
		chain 3 & 23.90 & 0.32 & 0.00 \\
		chain 4 & 7.75 & 0.87 & 0.00 \\
		chain 5 & 2.78 & 1.77 & 0.00 \\
		chain 6 & 1.12 & 3.30 & 0.00 \\
		chain 7 & 0.47 & 6.45 & 0.00 \\
		chain 8 & 0.23 & 12.65 & 0.01 \\
		chain 9 & 0.07 & 22.44 & 0.05 \\
		chain 10 & 0.02 & 0.00 & 0.23 \\
		chain 11 & 0.00 & 21.39 & 0.67 \\
		chain 12 & 0.00 & 13.92 & 1.52 \\
		chain 13 & 0.00 & 7.99 & 3.12 \\
		chain 14 & 0.00 & 4.29 & 5.46 \\
		chain 15 & 0.00 & 2.12 & 8.56 \\
		chain 16 & 0.00 & 1.11 & 12.18 \\
		chain 17 & 0.00 & 0.61 & 16.58 \\
		chain 18 & 0.00 & 0.34 & 22.37 \\
		chain 19 & 0.00 & 0.19 & 29.26 \\
		chain 20 & 0.00 & 0.13 & 0.00 \\
		\hline
		\end{tabular}
		\caption{Proportions (\%) of accepted equi-energy moves between chain 1 and other possible chains (mean on 100 runs of PTEEM). Idem for chains 10 and 20.}\label{tabEEmoves3}
		\end{center}
		\end{table}
		The mean acceptance rates of the equi-energy moves for PTEEM and of the exchange moves for PT were of 49\% and 61\% respectively.
		Note that we could implement the PT algorithm so that exchange moves can be proposed between any two chains and not only between adjacent chains. But in this case the mean acceptance rate of an exchange move would be much lower. In comparison the PTEEM algorithm has the advantage to propose exchanges moves between chains not necessarily adjacent, but more relevant in terms of energy levels.
		%In conclusion, in this example the PTEEM algorithm performs a better exploration of the parameter space than the PT algorithm, while in mean less exchanges between chains were performed for the PTEEM algorithm, compared to the PT algorithm.

\section{A complex problem: discovery of transcription factor binding motifs}\label{TFBS}
      \subsection{Model and  data}\label{ModelData}
\begin{figure}\label{WebLogo}
	  \begin{center}
		  \includegraphics[scale=0.6]{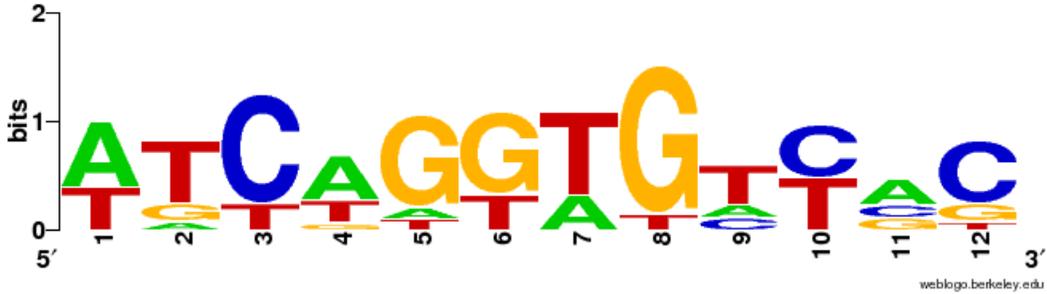}
	  \end{center}
	  \caption{WebLogo corresponding to the product multinomial used to generate data.}
	  \end{figure}

	  The  discovery of binding motifs in order to understand gene regulation is an important topic in biology. Indeed, a first step to understand gene expression is to know which are the corresponding binding sites of a common transcription factor (TFBS). The identification of these TFBS is a major computational problem, often studied these last twenty years (see for instance \cite{StormoHartzell}, \cite{LawrenceReilly}, \cite{LawrenceGS}, \cite{LiuNeuwaldLawrence} or \cite{Jensen2004}).\\
	  The data often consist of several homologous DNA sequences, and finding the TFBS is equivalent to identifying the starting positions of these sites in the sequences. Denote by $S$ the set of $M$ sequences, each one containing zero, one or more TFBS. Each sequence is made of four nucleotides: A, C, G or T. The TFBS are assumed to be of known length $w$. The length of the $m$th sequence is $L_m$, hence the number of possible starting positions for TFBS is denoted by $L_m^*= L_m -(w-1)$. The total number of motif sites is unknown and is denoted by $|A|$. As this number is unknown, the $M$ sequences (without their $w-1$ last nucleotides) are considered as one long sequence of length $L^*=\sum_{m=1}^M L_m^*$. This long sequence contains $|A|$ TFBS. To identify the most promising positions for the TFBS, we introduce a missing vector $A=(a_1,a_2\ldots,a_{L^*})$, where $a_i = 1$ if the $i$th position of the long sequence is the starting point of a TFBS, and $a_i= 0$ otherwise.  Given $A$, the set $S$ can be written as the union of two disjoint subsets: $S(A)\bigcup S(A^C)$, where $S(A)$ contains the aligned motifs of the identified TFBS, $S(A^C)$ representing the  background sequence. Two different models are used for these two subsets. Concerning the background sequence, the simplest model is a product multinomial model (see \cite{LiuNeuwaldLawrence}), but it has been shown that a Markov model is biologically more relevant and improves the results obtained (see \cite{Jensen2004}). However, it makes the motif discovery more difficult, as repeated patterns are local modes for the algorithms. Following \cite{KouZhouWong} we used a Markov model of order one based on the following transition matrix
	  $$\theta_0 =\left(
          \begin{array}{cccc}
            1 -3\alpha & \alpha &\alpha &\alpha\\
            \alpha & 1-3\alpha&\alpha&\alpha \\
            \alpha &\alpha&\alpha&1-3\alpha \\
          \end{array}
          \right) $$
	  where $\alpha = 0.12$. The parameter $\theta_0$ is assumed to be known (in practice it can be easily well estimated from the data). Concerning $S(A)$, it can be seen as a matrix of dimensions $|A| \times w$, with the BSFT in rows. The $k$th column contains the nucleotides in $k$th position of the $|A|$ sites. Let $C=(C_1, ..., C_w)$  be a count vector, where $C_k=(C_{kA},C_{kC},C_{kG},C_{kT})$ is the vector of the nucleotides counts in position $k$ of all the sites. The common pattern of the TFBS is modeled by a product multinomial distribution of parameter $\Theta = (\theta_1, ..., \theta_w)$  where $\theta_k = (\theta_{kA}, \theta_{kC}, \theta_{kG}, \theta_{kT})$ is a probability vector for the preference of the nucleotide types in position $k$. According to the model, each vector $C_k$  has a multinomial distribution with parameter $\theta_k$ independent of the other columns. For this example we used
	  \begin{displaymath}
	  \Theta = \left(\begin{array}{cccccccccccc}
		  0.6 & 0.1 & 0   & 0.6 & 0,1 & 0   & 0.3 & 0   & 0.2 & 0   & 0.5  & 0\\
		  0   & 0   & 0.8 & 0   & 0   & 0   & 0   & 0   & 0.2 & 0.5 & 0.25 & 0.7\\
		  0   & 0.2 & 0   & 0.1 & 0.8 & 0.7 & 0   & 0.9 & 0   & 0   & 0.25 & 0.2\\
		  0.4 & 0.7 & 0.2 & 0.3 & 0,1 & 0.3 & 0.7 & 0.1 & 0.6 & 0.5 & 0    & 0.1
	    \end{array}\right).
	  \end{displaymath}
	  The corresponding WebLogo (\cite{WebLogo}) is given in Figure \ref{WebLogo}.\\

	  To complete the model, conjugate prior distributions are considered. The distribution of $\Theta$  is  a product of Dirichlet with parameters $(\beta_1,\ldots, \beta_w)$: %, i.e. the $\theta_k$ are independent Dirichlet random variables with parameters $\beta_k$ of length 4.
	  \begin{displaymath}
	  \pi(\Theta) \propto \prod_{k=1}^w \theta_k^{\beta_k-1}, \qquad \textrm{where} \qquad \theta_k^{\beta_k}=\prod_{j=\{A,C,G,T\}} \theta_{kj}^{\beta_{kj}}.
	  \end{displaymath}
	  The prior probability of a component $a_i$ of $A$ is denoted by $p_0$, which is the "site abundance" parameter:
	  \begin{displaymath}
	    \pi(A \mid p_0) = p_0^{|A|} (1-p_0)^{L^*-|A|}.
	  \end{displaymath}
	  Finally, this parameter $p_0$ is assumed to follow a beta distribution $Be(a,b)$.
	%  \begin{displaymath}
	%    \pi(p_0) \propto p_0^{a-1} (1-p_0)^{b-1}.
	%  \end{displaymath}

	  From $\theta_0$ and $\Theta$, we generated $M=10$ background sequences of length $200$, and $|A|=20$ TFBS of length $w=12$. Two TFBS were introduced in each of the ten sequences, hence we obtained 10 sequences of length 224.

      \subsection{Classical approach: the Gibbs sampler}
	    To solve the challenging problem of identifying TFBS, bayesian approaches using Gibbs samplers were developed by \cite{LawrenceGS}, \cite{LiuNeuwaldLawrence}, or \cite{Liu}. The missing vector $A$ giving the starting positions of the TFBS is of interest, hence the aim is to build a Markov chain having the posterior distribution of $A$ as stationnary distribution.

	   % Noting that $S = S(A) \cup S(A^C)$, the complete-data likelihood of $(S, A)$ can be written as
	   % \begin{eqnarray*}
	   % \pi(S, A|\Theta, p_0) %&\propto& p(S|A,\Theta) p(A|p_0)\\
	   % &\propto& p(S(A)|A, \Theta) p(S(A^C)|A, \theta_0)p(A|p_0)\\
	   % &\propto& \frac{p(S(A)|A, \Theta)}{p(S(A)|A, \theta_0)}p(A|p_0)\\
	   % &\propto& \frac{1}{p(S(A)|A, \theta_0)}\prod_{k=1}^w \theta_k^{C_k} \times p_0^{|A|}(1-p_0)^{L^*-|A|}\\
	   % \end{eqnarray*}
	   % where $\theta_k^{C_k}=\prod_{j=\{A,C,G,T\}} \theta_{kj}^{C_{kj}}$.
%We note $p(S(A)|A, \theta_0)$ to make it clear that it is the probability of generating $S(A)$ from the background model, parametrized by $\theta_0$.\\
	   % The conditional posterior distribution is
	   % \begin{eqnarray*}
	   % \pi(\Theta, p_0| S, A) &\propto& P(S, A|\Theta, p_0) \pi(\Theta) \pi(p_0)\\
	   % &\propto& \frac{1}{p(S(A)|A, \theta_0)}\prod_{k=1}^w \theta_k^{C_k+\beta_k-1} \times p_0^{|A|+a-1}(1-p_0)^{L^*-|A|+b-1}.
	   % \end{eqnarray*}
	    %where $\theta_k^{C_k+\beta_k-1}=\prod_{j=\{A,C,G,T\}} \theta_{kj}^{C_{kj}+\beta_{kj}-1}$.
%The joint posterior distribution   $\pi(A, \Theta, p_0 \mid S)$ is proportional to $\pi(\Theta, p_0| S, A)$. However, we are interested only by the posterior distribution of $A$. Therefore

  Following \cite{KouZhouWong}, in order to obtain the posterior of interest $\pi(A \mid S)$, the collapsing technique of \cite{Liu} is used to integrate out the unknown parameters $\Theta$ and $p_0$ in the joint posterior distribution. As a consequence these parameters are not  updated at each iteration and the computation time is reduced. Moreover, the use of this technique facilitates the convergence of the Markov chain, as noted by \cite{Liu} and \cite{vanDyk}. The posterior of interest is given by
	   \begin{eqnarray}\label{postA}
	    \pi(A \mid S) &\propto& \frac{1}{\pi(S(A)|A, \theta_0)} \frac{\Gamma(|A|+a)\Gamma(L^*-|A|+b)}{\Gamma(L^*+a+b)} \prod_{k=1}^w \frac{\Gamma(C_k+\beta_k)}{\Gamma(|A|+|\beta_k|)},
	   \end{eqnarray}
	    with $\Gamma(C_k+\beta_k) = \prod_{j=\{A,C,G,T\}} \Gamma(C_{kj}+\beta_{kj})$.
	    Using  (\ref{postA}), a predictive update version of the Gibbs sampler has been proposed by \cite{LiuNeuwaldLawrence}. They suggest to update each component of $A$ independently of the others using the following predictive update formula:
	    \begin{eqnarray}\label{predictiveupdate}
	      \frac{p(a_i=1|A_{[-i]}, S)}{p(a_i=0|A_{[-i]}, S)} &=& \frac{1}{\pi(S(a_i) \mid A, \theta_0)} \times \frac{|A_{[-i]}|+a}{L^*-A_{[-i]}-1+b} \times \prod_{k=1}^w \Bigg(\frac{C_{k(-i)}+ \beta_k}{|A_{[-i]}|+|\beta_k|}\Bigg)^{C_{k(i)}},
	    \end{eqnarray}
	    where the following notations are used: $A_{[-i]}$ represents the vector $A$ without the $i$th component, $S(a_i)$ represents the sites of $A$ starting in position $i$,  $C_{k[-i]}=(C_{k[-i]A},C_{k[-i]C},C_{k[-i]G},C_{k[-i]T})$ is the vector of the nucleotides counts in position $k$ of all the sites, excluding the site starting in position $i$, $C_{k(i)}=(C_{k(i)A},C_{k(i)C},C_{k(i)G},C_{k(i)T})$ is the vector of the nucleotides counts in position $k$ of the site starting in position $i$ (this vector contains three 0 and one 1). We have $C_k= C_{k[-i]} + C_{k(i)}$.

      \subsection{EES algorithm}
	    The algorithms resulting from the Gibbs sampling approach, such as BioProspector (\cite{Bioprospector}) or AlignACE (\cite{AlignACE}), are often trapped into local modes and true motif patterns are not found. Therefore \cite{KouZhouWong} proposed to use the EES algorithm, which seems to improve the global TFBS search. In this algorithm, $K$ chains are used and the $l$th chain has the following target distribution
	    \begin{displaymath}
		\tilde{\pi}_l(A) \propto exp\Big(-\frac{h(A) \vee H_l}{T_l}\Big), \qquad \textrm{with} \qquad h(A)=-log(\pi(A \mid S)).
	    \end{displaymath}
For the target chain \cite{KouZhouWong} used a Gibbs sampler to generate the vector $A$. For the other chains, given the current sample $A$, they first estimated the common pattern  by a frequency counting. Then they built a new vector  according to the Bayes rule, which  is accepted according to a Metropolis-Hasting move (see \cite{KouZhouWong} for more details).

      \subsection{PTEEM algorithm}
	    In this algorithm, $N$ chains are used and the $l$th chain has the following target distribution
	    \begin{displaymath}
		\pi_l(A \mid S) = \pi(A \mid S)^{\frac{1}{T_l}},
	    \end{displaymath}
	    and are locally updated by Gibbs samplers. Concerning the first chain, the updating of each component of $A$ is done using the predictive update formula (\ref{predictiveupdate}). Concerning the $l$th chain ($l>1$), the predictive update formula to be used is the following:
	    \begin{eqnarray}\label{predictiveupdate2}
		\frac{p(a_i=1|A_{[-i]}, S)}{p(a_i=0|A_{[-i]}, S)} &=& \Bigg(\frac{1}{\pi(S(a_i) \mid A, \theta_0)} \times \frac{|A_{[-i]}|+a}{L^*-A_{[-i]}-1+b} \times \prod_{k=1}^w \Bigg(\frac{C_{k[-i]}+ \beta_k}{|A_{[-i]}|+|\beta_k|}\Bigg)^{C_{k(i)}}\Bigg)^{\frac{1}{T_l}}
	    \end{eqnarray}
	    Concerning a proposed equi-energy move between two chains  $l_1$ and $l_2$ of current states $A_{l_1}$ and $A_{l_2}$, the acceptance probability is given by:
	    \begin{displaymath}
	    \rho= \min\{1, \frac{\pi_{l_1}(A_{l_2} \mid S)\pi_{l_2}(A_{l_1} \mid S)}{\pi_{l_1}(A_{l_1} \mid S)\pi_{l_2}(A_{l_2} \mid S)}\},
	    \end{displaymath}
	    with
	    \begin{eqnarray*}
	    \frac{\pi_{l_1}(A_{l_2} \mid S)\pi_{l_2}(A_{l_1} \mid S)}{\pi_{l_1}(A_{l_1} \mid S)\pi_{l_2}(A_{l_2} \mid S)} &=&  \Big(\frac{\pi(A_{l_2} \mid S)}{\pi(A_{l_1} \mid S)}\Big)^{\frac{1}{T_{l_1}}-\frac{1}{T_{l_2}}}\\
	      &=& \Bigg[ \frac{\pi(S(A_{l_1}) \mid A_{l_1}, \theta_0)}{\pi(S(A_{l_2}) \mid A_{l_2}, \theta_0)} \times \frac{\mathcal{B}(|A_{l_2}|+a,L^*-|A_{l_2}|+b)}{\mathcal{B}(|A_{l_1}|+a,L^*-|A_{l_1}|+b)} \\
	      & & \times \prod_{k=1}^w \frac{\Gamma(C_{l_2 k}+\beta_k)}{\Gamma(C_{l_1 k}+\beta_k)} \times \prod_{k=1}^w \frac{\Gamma(|A_{l_1}|+|\beta_k|)}{\Gamma(|A_{l_2}|+|\beta_k|)} \Bigg]^{\frac{1}{T_{l_1}}-\frac{1}{T_{l_2}}}.
	    \end{eqnarray*}

      \subsection{Results}
	  The algorithms EES and PTEEM as explained above were run 10 times each, on the data presented in \ref{ModelData}. For each run of PTEEM, $N=15$ chains were used, with a burn-in of 200 iterations and a post-burn-in of 800 iterations, resulting in 15000 local moves and 1000 proposed global moves. For each run of EES, $K=9$ chains were used, with $p_{ee}=0.1$, a burn-in of 200 iterations and a post-burn-in of 800 iterations, among which 100 iterations were used to construct energy rings. It results in 18160 local moves and approximately 1640 global moves.

	  \paragraph{Calibration}~~\\
	      Concerning PTEEM, 5 energy rings were used, with energy levels regularly spaced on a logarithmic scale between 10 and 100, giving levels 10, 17, 78, 31.62, 56.23 and 100. The temperatures have their inverses regularly spaced between 1 and 1/1.3, giving $T_{min}=1$ and $T_{max}=1.3$.
	      Concerning EES, 9 energy rings were used, with energy levels regularly spaced on a logarithmic scale between 10 and 100. The temperatures used are the following: 1, 1.001, 1.002, 1.005, 1.01, 1.02, 1.06, 1.1 and 1.3. This choice has been made in order to permit equi-energy jumps between chains. For instance, fixing $T_1=1$ and $T_2=1.1$, no jumps would have been possible between the first and the second chains, because the energies of chains associated with these temperatures are too different.
	
	  \paragraph{Local and global moves}~~\\
	      Concerning the 10 runs of PTEEM, 55.71\% of the proposed equi-energy moves were accepted, allowing a good mixing of the chains. The first chain exchanged states relatively easily with chains of lower orders, and it exchanged states even with chains 10 or 11.
	      Concerning the 10 runs of EES, the last 8 chains were locally updated by Metropolis-Hastings algorithms: approximately 15\% of new states proposed for the second to fifth chains were accepted, but only 2.9\% were accepted for chain 8 and 0.7\% for chain 9. The proposed equi-energy jumps were mainly accepted (86\% in mean), but it is noticeable that very few jumps were proposed between the first and the second chain (26 jumps in mean during the 1000 iterations). Indeed, the states of these two chains often had energies quite different. As an example, the second chain never obtained a state in the first energy ring.

\begin{figure}[H]
	      \begin{center}
		      \includegraphics[scale=0.4]{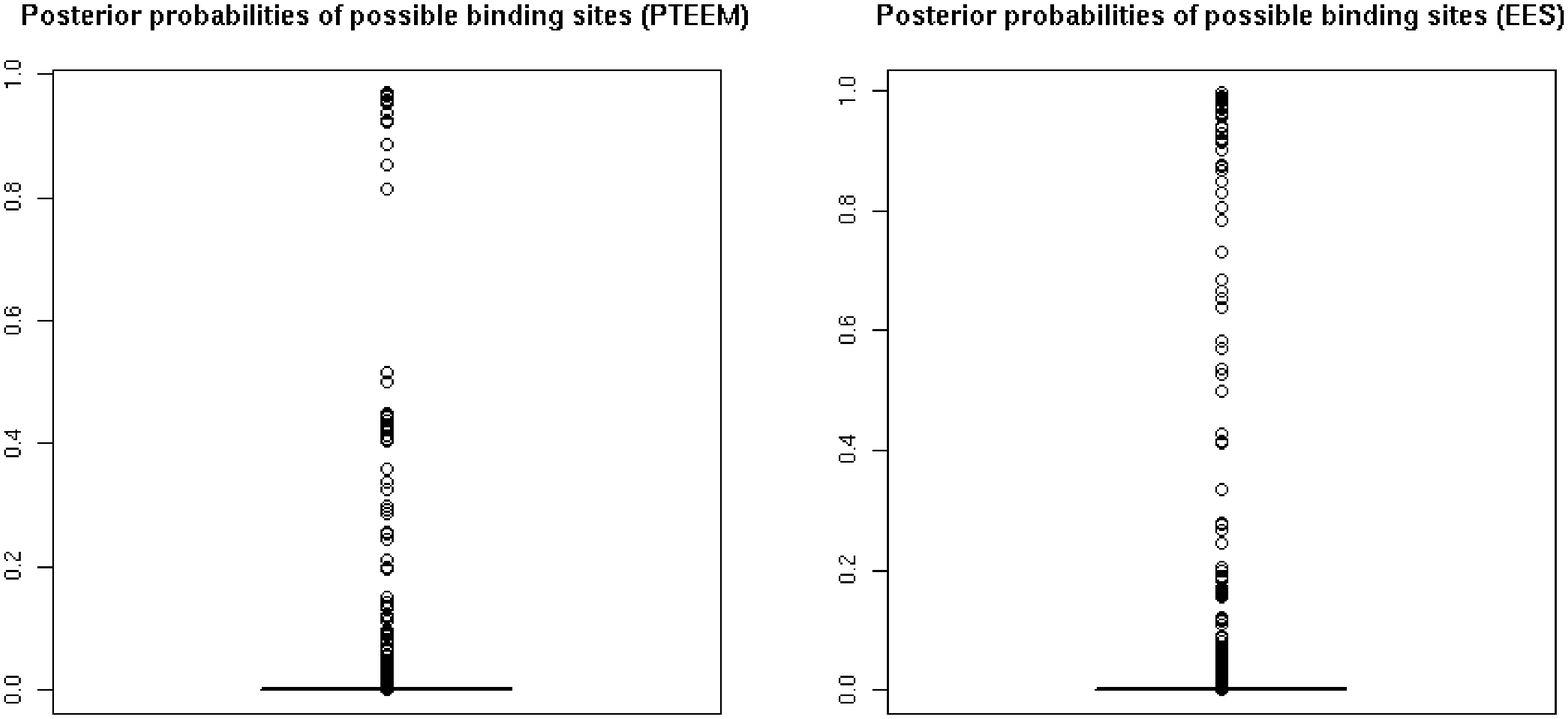}
	      \end{center}
	      \caption{representing posterior probabilities  $P(a_i=1 \mid S), i=1,\ldots,L^*$, obtained during a run of PTEEM, and during a run of EES.}
\label{ProbPost_TFBS}	      
\end{figure}

	  \paragraph{Identification of the TFBS}~~\\
	      Results of the 10 runs of PTEEM were quite similar, as opposition to the 10 runs of EES. Figure \ref{ProbPost_TFBS} represents two boxplots representing empirical posterior probabilities  $P(a_i=1 \mid S), i=1,\ldots,L^*$ obtained during a run of PTEEM, and during a run of EES. Hence these boxplots represent the posterior probabilities of each possible position to be the starting point of a TFBS.
	      Only the positions associated to high posterior probabilities are relevant. On the boxplots of figure \ref{ProbPost_TFBS} for instance, we decided to keep only the positions with posterior probabilities higher than 0.8. Concerning the 10 runs of PTEEM, they identified 16 sites among 20. Among them, 15 were identified with exactly the true starting positions, and 1 was identified with 3 other positions (positions 877, 880 and 883 were kept, the true one being 880).
	      Concerning the 10 runs of EES, they identified in mean 15.6 sites among 20. Among them, 9.6 were identified with exactly the true starting positions, and 6 were identified with phase-shifted positions or several positions. For example, a site has been identified by positions 1784 and 1791 while the true one was 1784, and another has been identified by position 27 while the true one was 26. Notice that 5 EES runs among 10  obtained similar results as the PTEEM runs.

	  \paragraph{Conclusion on these results}~~\\
	      Results obtained by EES could be improved with a better calibration. In particular, using more chains would improve the results (with a supplementary computational cost). However, the low number of jumps proposed between chains 1 and 2 is noticeable. It could be due to temperatures too far from each others, but as we used $T_1=1$ and $T_2=1.001$, it should be due to the Metropolis-Hastings algorithm used to update the second chain. This algorithm could have difficulties to propose relevant states. Indeed, it is not easy to find a good proposal law for new states, and maybe the method proposed by \cite{KouZhouWong} is not the best possible. The difficulties encountered to calibrate the EES and the associated Metropolis-Hastings algorithms are a disadvantage to the use of this algorithm. In comparison, the calibration of PTEEM is much easier. Indeed, the Gibbs sampler does not need to be calibrated, and if temperatures and energy levels are well chosen, the number of accepted equi-energy moves between chains is sufficiently large to allow good mixing of the chains. We did not encounter difficulties to calibrate these parameters, suggestions of \ref{ChoiceTempEnergy} giving good results.\\
	      Concerning the results obtained on this challenging example, those obtained with PTEEM were slightly better than those obtained with EES. The mixing of the chain of interest was more efficient in PTEEM. Hence, PTEEM identified exactly most of the true starting positions of TFBS, while EES tended to identified TFBS with several positions or phase-shifted positions. That means that EES was most often trapped in local phase-shift modes.\\
	      Note that this phase-shift problem is encountered by most of the methods used to identify TFBS, and solutions have been proposed, see \cite{Liu} or \cite{LawrenceGS}. Implementation of these solutions in the algorithms PTEEM or EES is absolutely possible. Similarly, improvements can be carried out to these algorithms to allow TFBS of unknown length, several motifs of TFBS, or TFBS made of several non contiguous blocks, see  \cite{Jensen2004} and \cite{Bioprospector}. % However, it would be easier to implement these improvements in a Gibbs sampler, and hence in PTEEM, than in EES.

\section{Discussion}

      In this paper a new algorithm combining Parallel Tempering and Equi-Energy Sampler was proposed. Inspired by the original  idea of the EES, it is based on the use of energy rings. 
      Thanks to relevant equi-energy moves, the proposed PTEEM algorithm allows a good exploration of the parameter space and good mixing of the generated Markov chains, while ensuring the reversibility of the exchange moves. Therefore the generated Markov process theoretically converges to $\pi^*$, and the first chain generates samples corresponding to the distribution of interest $\pi$.

      Compared to PT, this new algorithm has the same theoretical properties, while outperforming it. The drawback is that an energy ladder is needed, but we explained simple and practical ways to obtain a relevant ladder, which proved to be efficient.

      Compared to the original EES, this new algorithm has the advantage to be based on Monte Carlo Markov chains theory, which is quite simple to use and to understand, even for non-experimented users. Moreover, the asymptotic variance of PTEEM is smaller or equal to those of the EES, as noted by \cite{Atchade2010b} which compared MCMC algorithms and adaptive MCMC algorithms like the EES.
      On a practical point of view, the PTEEM needs less storage than the EES, since all iterations from the past are not kept in memory. Moreover, it can be coupled with a Gibbs sampler, unlike the original EES because of an energy truncation. However, the  EES could be modified to be used without energy truncation.
      %Now it is known that usually when a Gibbs sampler can be used, it gives better results than a Metropolis-Hastings. All the more that the performances of a Metropolis-Hastings algorithm depends on the proposal distribution, whose choice can be non trivial, see Section \ref{TFBS}.
      %In the cases where a Gibbs sampler can be used, our feeling is that this new algorithm can give results at least equivalent to those obtained with EES.
      On our examples, PTEEM gave results at least as good as those obtained with an EES.

      A direction for future research is to investigate further the theoretical properties of the PTEEM algorithm, by comparing convergence rates of PTEEM and PT algorithms for instance. Besides an automatic way to build the energy rings could be inspired from the approach of \cite{ZhouWong2008} to reconstruct the energy landscape of the target density. %A run of PTEEM is make with a large number of rings and the distribution of samples in the rings is observed to deduce the larger and lower values of the energy. Then there are two possibilities: either the obtained distribution is studied and some rings are grouped to get some relevant rings, i.e. having about the same number of simulations, else a tree of sublevel sets is built as in Zhou and Wong (2008) and at least one ring for each local minimum of energy and one ring for each local maximum of energy are chosen. It is probably longer than a well calibrated run of PTEEM. %this rule, more complicated, \\
      Finally, an adaptive PTEEM algorithm to finely tune the temperatures and/or the energy levels during a run would also be of interest.

\bibliography{references}

\appendix

\section{Formula used for the comparisons in case of a Gibbs sampler}\label{formulesex}
	\subsection{Joint posterior densities}
	Write $x=(\mu,\sigma^{-2},w,c,\beta )$.
	The joint posterior density from which the parameters should be drawn is:
	\begin{eqnarray*}
	\pi(x)%=p(\mu,\sigma^{-2},w,c,\beta \mid y) &\propto& p(y \mid \mu,\sigma^{-2},c,\beta,w) p(\mu,\sigma^{-2},c,\beta,w),\\
				&\propto& p(y \mid \mu,\sigma^{-2},c) p(\mu,\sigma^{-2},c,\beta,w).
	\end{eqnarray*}
	Hence the $i${th} chain should be drawn from
	\begin{displaymath}
	\pi_i(x) \propto \pi(x)^{\frac{1}{T_i}} \propto p(y \mid \mu,\sigma^{-2},c)^{\frac{1}{T_i}} p(\mu,\sigma^{-2},w,c,\beta)^{\frac{1}{T_i}}.
	\end{displaymath}
	However, as noted by \cite{JasraStephensHolmes2007b} and \cite{Behrens}, tempering the whole posterior is problematic as there is no guarantee that the tempered posterior will remain proper. As a consequence, only the likelihood contribution is tempered and the priors are left untempered. The $i${th} chain is then drawn from
	\begin{eqnarray*}
	\pi'_i(x) &\propto& p(y \mid x)^{\frac{1}{T_i}} p(x).%,\\
		%&\propto& p(y \mid \mu,\sigma^{-2},c)^{\frac{1}{T_i}} p(\mu \mid \xi,\kappa^{-1})p(\sigma^{-2} \mid \alpha,\beta)p(c \mid w)p(w \mid \delta)p(\beta \mid g,h).
	\end{eqnarray*}
%We have
%	\begin{eqnarray*}
%	\pi'_i(x) %&\propto& p(y \mid \mu,\sigma^{-2},c)^{\frac{1}{T_i}} p(\mu \mid \xi,\kappa^{-1})p(\sigma^{-2} \mid \alpha,\beta)p(c \mid w)p(w \mid \delta)p(\beta \mid g,h)\\
	%&\propto& \Big[\prod_{p=1}^k (\sigma_p \sqrt(2\pi))^{-m_p} exp\Big(-\sum_{l=1}^n \frac{(y_l-\mu_{c_l})^2}{2\sigma_{c_l}^2}\Big)\Big]^{\frac{1}{T_i}} \Big[\prod_{p=1}^k \frac{\kappa^{\frac12}}{\sqrt{2\pi}} exp\Big(-\frac12 (\mu_p - \xi)^2 \kappa\Big)\Big]\\
	%& & \times \Big[\prod_{p=1}^k \frac{\beta^{\alpha}}{\Gamma(\alpha)}\sigma_p^{-2(\alpha-1)} exp(-\beta \sigma_p^{-2})\Big]\Big[\frac{n!}{\prod_{p=1}^k m_p!}\prod_{p=1}^k w_p^{m_p}\Big]\\
%	& & \times \Big[\frac{1}{B(\delta,\ldots,\delta)}\prod_{p=1}^k w_p^{\delta-1}\Big]\Big[\beta^{g-1}exp(-h\beta)\frac{h^g}{\Gamma(g)}\Big],
%	\end{eqnarray*}
%	where
%	\begin{displaymath}
%	B(\delta,\ldots,\delta)=\frac{\prod_{p=1}^k \Gamma(\delta)}{\Gamma(\sum_{p=1}^k \delta)}= \frac{\Gamma(\delta)^k}{\Gamma(k\delta)}.
%	\end{displaymath}
	
	\subsection{Full conditional distributions}~~\\
	Concerning the $i${th} chain, the full conditional distributions to be used in the Gibbs sampler of the algorithms are easily obtained through conjugacy. We use the following notations:
	\begin{eqnarray*}
	x_i = (\mu_i,\sigma_i^{-2},w_i,c_i,\beta_i),
	&& \mu_i = (\mu_{i1},\mu_{i2},\ldots,\mu_{ik}),\\
	\sigma_i^{-2} = (\sigma_{i1}^{-2},\sigma_{i2}^{-2},\ldots,\sigma_{ik}^{-2}),
	&&w_i =(w_{i1},w_{i2},\ldots,w_{ik}),\\
	c_i = (c_{i1},c_{i2},\ldots,c_{in}),
	&&m_i = (m_{i1},m_{i2},\ldots,m_{ik}),
	\end{eqnarray*}
%	\vspace{0.2cm}
with $p = 1,\ldots,k$ index of component and
	$l = 1,\ldots,n$ index of observation.
	For $\mu_i$, $\sigma^{-2}_i$ and $w_i$ the full conditional distributions are the following
	\begin{displaymath}
		\mu_{ip} \mid \sigma_{ip}^{-2},y,c_i,\xi,\kappa^{-1} \sim \mathcal{N}\Bigg(\Big(\frac{m_{ip} \sigma^{-2}_{ip}}{T_i}+\kappa\Big)^{-1}\Big(\frac{\sigma^{-2}_{ip}}{T_i}\sum_{l : c_{il}=p}y_l +\xi \kappa\Big),\Big(\frac{m_{ip} \sigma^{-2}_{ip}}{T_i}+\kappa\Big)^{-1}\Bigg),
	\end{displaymath}
	\begin{displaymath}
		\sigma_{ip}^{-2} \mid \mu_{ip},y,c_i,\alpha,\beta_i \sim \Gamma\Big(\alpha+\frac{m_{ip}}{2T_i},\beta_i+\sum_{l : c_{il}=p} \frac{(y_l-\mu_{ip})^2}{2T_i}\Big),
	\end{displaymath}
	\begin{displaymath}
		w_i \mid c_i,\delta \sim D(\delta+m_{i1},\delta+m_{i2},\ldots,\delta+m_{ik}).
	\end{displaymath}
	For the allocation vector $c$, the full conditional distribution is multinomial with the following probabilities:
	\begin{displaymath}
	p_i(c_{il}=p \mid y,\mu_i,\sigma_i^{-2},w_i) \propto \frac{1}{\sigma_{ip}^{\frac{1}{T_i}}}exp\Big(-\frac{(y_l-\mu_{ip})^2}{2\sigma_{ip}^2 T_i}\Big) w_{ip}.
	\end{displaymath}
	The parameter $\beta_i$ has the following full conditional distribution:
	\begin{displaymath}
	\beta_i \mid \sigma_i^{-2},\alpha,g,h \sim \Gamma\Big(g+k\alpha,h+\sum_{p=1}^{k} \sigma_{ip}^{-2} \Big).
	\end{displaymath}

	\subsection{Acceptance rate of an equi-energy move}~~\\
	Assuming that two chains $i$ and $j$ are selected from an energy ring to be swapped, the acceptance probability of  an equi-energy move proposed between two chains is given by
	\begin{displaymath}
		\min\left(1,  \frac{\pi'_i(x_j)\pi'_j(x_i)}{\pi'_i(x_i)\pi'_j(x_j)}\right)
=
		\min\left(1,  \Big(\frac{p(y|x_i)}{p(y|x_j)}\Big)^{(1/T_j-1/T_i)}\right),
	\end{displaymath}
where
\be*
p(y|x) & \propto &
\prod_{p=1}^k (\sigma_p \sqrt(2\pi))^{-m_p} exp\Big(-\sum_{l=1}^n \frac{(y_l-\mu_{c_l})^2}{2\sigma_{c_l}^2}\Big).
\e*
%	with
%	\be*
%	\frac{\pi_i'(x_j)\pi_j'(x_i)}{\pi_i'(x_i)\pi_j'(x_j)} = \Bigg[\frac{\prod_{p=1}^k \sigma_{jp}^{-m_{jp}}}{\prod_{p=1}^k \sigma_{ip}^{-m_{ip}}}\Bigg]^{\frac{1}{T_i}-\frac{1}{T_j}} exp\Bigg[-\frac12 \Big(\frac{1}{T_i}-\frac{1}{T_j}\Big) \Big(\sum_{l=1}^n (y_l-\mu_{jc_{jl}})^2\sigma_{jc_{jl}}^{-2} - \sum_{l=1}^n (y_l-\mu_{ic_{il}})^2\sigma_{ic_{il}}^{-2}\Big)\Bigg]
%	\e*

\section{Proofs of Proposition \ref{prop1} and Lemma \ref{lemme1}}
	\subsection{Proof of Proposition \ref{prop1}}\label{proofprop1}
		During an iteration of the PTEEM algorithm all chains are locally updated by a MCMC algorithm and an exchange move is proposed. By assumption,  $PL_i(.,.)$ is reversible with stationary distribution $\pi_i$. It is then clear that $PL=\prod_{i=1}^N PL_i$ is also reversible. Let $A \in {\cal B}({\cal X})^N$, which can be written as $A_1 \times A_2 \times \ldots \times A_N$, with $A_i \in \cal X$. We have
			    \begin{eqnarray*}
			    \pi^*(A)
				    &=& \int_{{\cal X}^N} PL(s,A)\pi^*(ds),
			    \end{eqnarray*}
			    which implies that  $\pi^*$ is the stationary distribution of $PL(.,.)$.
			    Then, the transition kernel $PE$ can be written as
			    \begin{equation}\label{noyauPE}
			    PE(s,s') = q(s,s')\rho(s,s')+ \int_{\cal X} q(s,s'')(1-\rho(s,ds'')) \mathds{1}_{\{s'\}}(s).
			    \end{equation}
			    A sufficient condition to satisfy the detailed balance condition is the following:
			    \begin{equation}\label{intEquals}
			    q(s,ds')\rho(s,s')\pi^*(ds) = q(s',ds)\rho(s',s)\pi^*(ds').
			    \end{equation}
		In the PTEEM algorithm, the two candidate chains to exchange their actual states are chosen uniformly among all chains in the same energy ring. Hence we have $q(s,s')=q(s',s)$. Using (\ref{probaccep}), it follows that  (\ref{intEquals}) is satisfied, and the detailed balance condition  holds. Therefore the transition kernel $PE$ for the equi-energy move is reversible, with stationary distribution $\pi^*$.
		The transition kernels $PE$ and $PL$ are reversible with stationary distribution $\pi^*$. It is then clear that $P$ is also reversible and that $\pi^*$ is its stationary distribution.  	
		In addition, each $PL_i$ is supposed to be strongly $\lambda$-irreducible and  aperiodic, hence $PL$ is aperiodic and strongly $\lambda_N$-irreducible. Since  $PE$ is just an exchange kernel between two actual states it is clear that $P=PE * PL$ is also strongly $\lambda_N$-irreducible and aperiodic.
		Theorem 1 of \cite{Tierney94} then allows to conclude.
		\cqfd	An automatic way to build the energy rings could be inspired from the approach of \cite{ZhouWong2008} to reconstruct the energy landscape of the target density. A run of PTEEM is make with a large number of rings and the distribution of samples in the rings is observed to deduce the larger and lower values of the energy. Then there are two possibilities: either the obtained distribution is studied and some rings are grouped to get some relevant rings, i.e. having about the same number of simulations, else a tree of sublevel sets is built as in Zhou and Wong (2008) and at least one ring for each local minimum of energy and one ring for each local maximum of energy are chosen. It is probably longer than a well calibrated run of PTEEM. %this rule, more complicated, \\

	\subsection{Proof of Lemma \ref{lemme1}}\label{prooflemme1}
		From Proposition \ref{prop1}, $S$ is reversible with stationary distribution $\pi^*$, and strongly $\lambda_N$-irreducible. It follows that $S$ is positive.
		Note that a state $s'$ reached from a starting point $s$ after an iteration of PTEEM can not be part of a set $A \in \mathcal{X}^N$ such that $\pi^*(A)=0$ (proof inspired from \cite{RobertsRosenthal2006}, Theorem 8).\\
% 		Indeed, at each iteration either a move is perfomed by the sampler, or the sampler stays in the same state. The kernel $P$ can be written, $\forall A \in \mathcal{X}^N$
% 		\begin{equation}\label{eq1lemme}
% 		 P(s,A)=r(s)M(s,A)+(1-r(s))\mathbb{1}_{s}(A),
% 		\end{equation}
% 		with $r(s)$ the probability to move from $s$ during an iteration. $M$ is the kernel conditional on moving. The probability measure $M(s,.)$ is absolutely continuous with respect to $\lambda_N(.)$ for all $s \in \mathcal{X}^N$. Let $A$ such that $\pi^*(A)=1$ and $\pi^*(A^C)=0$. From the positivity of $\pi^*$, we have $\lambda_N(A^C)=0$. By absolute continuity, we have $M(s,A^C)=0$ and $M(s,A)=1$. Hence if the sampler moves from any starting state $s$, it will necessarily move into $A$.\\

		To show that $S$ is Harris-recurrent we use Theorem 2 of \cite{Tierney94} that characterizes Harris-recurrent chains as follows:
		a Markov chain is Harris-recurrent if and only if the only bounded functions $h$ satisfying
		\begin{equation}\label{harmonic}
		\mathbb E (h(S^{(n)})|s_0)=\mathbb E (h(S^{(1)})|s_0)=h(s_0), \qquad \forall n \in \mathbb{N},
		\end{equation}
		are the constant functions. Functions $h$ satisfying (\ref{harmonic}) are called harmonic.
		We use Theorem 6.80 of \cite{MonteCarloStatMethods}, inspired from \cite{Athreya96} as follows:\\
		If the transition kernel $P$ satisfies: $\exists B \in \mathcal B{({\cal X})}^N$ such that
		\begin{enumerate}
		\item[(i)] $\forall s_0$, $\sum_{n=1}^{\infty} \int_{B} P^n(s_0,s)d\mu(s) >0$, with $\mu$ the initial distribution of the chain.%, $S_0 \sim \mu$.
		\item[(ii)] $\inf_{s,s' \in B} P(s,s')>0$
		\end{enumerate}
		Then, for $\pi^*$-almost all $s_0$,
		\begin{equation}\label{lim}
		 \lim_{n\rightarrow \infty} \sup_{A \in \mathcal B{({\cal X})}^N} \Big|\int_{A} P^n(s_0,s)ds - \int_{A}\pi^*(s)ds\big|=0,
		\end{equation}
%		The larger the set $B$, the easier it is to verify (i) and the harder it is to verify (ii). These two assumptions (i) et (ii) imply that the chain is irreducible and aperiodic.\\
		To apply this result, notice that Assumptions (i) and (ii) are verified for $B=\mathcal{X}^N$. Equation (\ref{lim}) is then satisfied for $\pi^*$-almost all $s_0$.\\	
		%Indeed, $S$ is strongly $\lambda_N$-irreducible, hence $\forall (s,s') \in \mathcal{X}^N \times \mathcal{X}^N$, $P(s,s')>0$, and (ii) is satisfied on $\mathcal{X}^N$. Similarly, $\mathcal{X}^N$ is accessible from any initial state $s_0$, and (i) is satisfied.\\
	
		Using
		\begin{displaymath}
		 \| \mu\|_{TV}  = \sup_{A \in \mathcal{B ({\cal X})}^N} |\mu(A)| = \frac12 \sup_{|h|<1} |\int h(x) \mu(dx) |,
		\end{displaymath}
		this equation (\ref{lim}) can be written as
% 		\begin{eqnarray*}
% 		\sup_{A \in \mathcal B{({\cal X})}^N} \Big|\int_{A} P^n(s_0,s)ds - \int_{A}\pi^*(s)ds\big| &=&  \sup_{A \in \mathcal B{({\cal X})}^N} \Big|\int_{A} \big(P^n(s_0,s)ds - \pi^*(s)\big)ds\big| \\
% 		&=& \frac12 \sup_{|h|<1} \Big|\int h(s) \big(P^n(s_0,s)ds - \pi^*(s)\big)ds\Big|\\
% 		&=& \frac12 \sup_{|h|<1} \Big|\int h(s) P^n(s_0,s)ds - \int h(s)\pi^*(s)ds\Big|\\
% 		&=& \frac12 \sup_{|h|<1} \Big|E[h(S_n)|s_0]-E_{\pi^*}[h(s)]\Big|
% 		\end{eqnarray*}
% 		Hence the result of the theorem 6.80 of Robert and Casella can be written as:
		\begin{displaymath}
		 \lim_{n\rightarrow \infty} \sup_{|h|<1} \Big|E[h(S_n)|s_0]-E_{\pi^*}[h(s)]\Big| =0.
		\end{displaymath}
		We can extend this result for all bounded function $h$.
%		\begin{displaymath}
%		 \lim_{n\rightarrow \infty} \sup \Big|E[h(S_n)|s_0]-E_{\pi^*}[h(s)]\Big| =0
%		\end{displaymath}
		Moreover, if $h$ bounded satisfies (\ref{harmonic}), then $E[h(S_n)|s_0]=h(s_0)$.
		We then have $h(s_0)=E_{\pi^*}[h(s)]$ for $\pi^*$-almost all $s_0$, and $h$ is $\pi^*$-almost everywhere constant and equal to $\mathbb{E}_{\pi^*}(h(S))$. Analysis similar to that in the proof of Theorem 6.80 of \cite{MonteCarloStatMethods} shows that $h$ is everywhere constant and equal to $\mathbb{E}_{\pi^*}(h(S))$.
	       The Harris-recurrence then follows.
	\cqfd

\end{document}